\documentclass[11pt]{article}
\pdfoutput=1 
\usepackage{amssymb}
\usepackage{amsmath}
\usepackage{amstext}
\usepackage{lipsum}
\usepackage{graphicx,epsfig}
\usepackage{epsfig}
\usepackage{verbatim} 
\usepackage{fancybox}
\usepackage{color}
\usepackage{ulem}
\usepackage{enumitem}
\usepackage{subfigure}
\usepackage{bbm}
\usepackage{parskip}
\usepackage{dsfont}
\usepackage[numbers,sort&compress]{natbib}
\allowdisplaybreaks

\newcommand{\Comment}[1]{{}}
\definecolor{darkblue}{rgb}{0.15,0.35,0.55}
\definecolor{comment}{rgb}{1,0.4,0.4}
\definecolor{reddish}{rgb}{0.65, 0.2, 0.2}
\usepackage[linktocpage=true]{hyperref}
\hypersetup{
colorlinks=true,
citecolor=darkblue,
linkcolor=reddish,
urlcolor=darkblue,
pdfauthor={},
pdftitle={},
pdfsubject={}
}

\newcommand{\be}{\begin{equation} }
\newcommand{\ee}{\end{equation} }
\newcommand{\bea}{\begin{eqnarray}}
\newcommand{\eea}{\end{eqnarray}}
\newcommand{\beas}{\begin{eqnarray*}}
\newcommand{\eeas}{\end{eqnarray*}}

\def\({\left(}
\def\){\right)}

\newcommand{\rd}{{\rm d}}

\newcommand{\delp}{\pi}

\newcommand{\Mpl}{M_{\rm Pl}}

\def\gsim{ \lower .75ex \hbox{$\sim$} \llap{\raise .27ex \hbox{$>$}} }
\def\lsim{ \lower .75ex \hbox{$\sim$} \llap{\raise .27ex \hbox{$<$}} }

\def\xyma{\xymatrix@M.7em}
\def\xymas{\xymatrix@M.1em}
\newcommand{\ba}{\begin{eqnarray}}
\newcommand{\ea}{\end{eqnarray}}

\definecolor{darkred}{rgb}{0.7,0.3,0.3}
\definecolor{darkgreen}{rgb}{0.2,0.7,0.3}
\definecolor{greyish}{rgb}{.90,.90,.90}
\definecolor{greyish2}{rgb}{.96,.96,.96}
\definecolor{darkblue2}{rgb}{0.3,0.4,0.9}
\usepackage{pifont}
\usepackage{xcolor,colortbl}

\usepackage{tcolorbox}

\title{}
\author{}

\numberwithin{equation}{section}

\begin{document}

\setcounter{page}{1}
\renewcommand{\thefootnote}{\fnsymbol{footnote}}
~
\vspace{.80truecm}
\begin{center}
{\LARGE \bf{Existence and Instability of Novel Hairy Black Holes in Shift-symmetric Horndeski Theories}}\\ \vspace{.2cm}
\end{center}

\vspace{1truecm}
\thispagestyle{empty}
\centerline{{\large Justin Khoury,\footnote{\href{jkhoury@sas.upenn.edu}{\texttt{jkhoury AT sas.upenn.edu}}}
Mark Trodden,\footnote{\href{mailto:trodden@physics.upenn.edu}{\texttt{trodden AT physics.upenn.edu}}}
and Sam S. C. Wong\footnote{\href{mailto:scswong@sas.upenn.edu}{\texttt{scswong AT sas.upenn.edu}}}}}
\vspace{.5cm}
 
 \centerline{{\it Center for Particle Cosmology, Department of Physics and Astronomy,}}
 \centerline{{\it University of Pennsylvania, 209 South 33rd St., Philadelphia, PA 19104}} 
 \vspace{.25cm}

 \vspace{.8cm}
\begin{abstract}
\noindent
Shift-symmetric Horndeski theories admit an interesting class of Schwarzschild black hole solutions exhibiting time-dependent scalar hair. By making use of Lema\^{i}tre coordinates, we analyze perturbations around these types of black holes, and demonstrate that scalar perturbations around black hole backgrounds inevitably have gradient instabilities. Taken together with previously established results, this newly-discovered instability rules out black holes with time-dependent scalar hair in Horndeski theories.
\end{abstract}

\newpage

\setcounter{tocdepth}{2}
\tableofcontents

\renewcommand*{\thefootnote}{\arabic{footnote}}
\setcounter{footnote}{0}

\section{Introduction}
The past decade has seen stunning successes of observational efforts to elucidate the properties of astrophysical black holes, with two of the most high-profile results being the measurement of gravitational waves by the LIGO collaboration~\cite{TheLIGOScientific:2016pea}, and the imaging of super massive black hole shadows by the Event Horizon Telescope~\cite{Akiyama:2019cqa}. These remarkable achievements have paved the way for a new era of black hole science. One way to interpret these results is as a direct vindication of well-known predictions of General Relativity (GR). However, as is frequently the case with breakthrough observations, it is also possible to turn these measurements into new precision tools with which to probe theoretical ideas that go beyond our already established theories. In particular, there exist candidate theories of modified gravity that are consistent with existing observations, and some of these are of interest as possible explanations for a number of unexplained cosmological observations. As black hole observations offer increasing precision and statistics, they provide an invaluable tool with which to constrain, rule out, or perhaps find support for some of these theoretical constructions. One example of such an observable is the tails of gravitational waves from the ringdown phase of black hole mergers. The ringdown phase is characterized by quasinormal modes of the system~\cite{Berti:2009kk} which are sensitive to new operators in modified gravity. Hairy black hole solutions in some modified gravity theories are known to give rise to quasinormal modes that differ from the predictions of GR, and hence that are amenable to testing in this way. 

One reasonably general class of modified gravity theories can be obtained by considering all scalar-tensor operators (those involving just the metric and a real scalar field) that give rise to second order equations of motion (and hence avoid the existence of ``Ostrogradsky" ghost instabilities~\cite{Woodard:2015zca}). Models captured by this framework are known as {\it Horndeski} theories~\cite{Horndeski:1974wa}, which can be shown to be equivalent to generalized Galileon theories in curved space-time~\cite{Deffayet:2009mn,Deffayet:2011gz,Kobayashi:2011nu}. 
This class of theories has been further enlarged with Gleyzes-Langlois-Piazza-Vernizzi (GLPV)~\cite{Gleyzes:2014dya} and Degenerate Higher Order Scalar-Tensor (DHOST)~\cite{Langlois:2015cwa,Crisostomi:2016czh,BenAchour:2016fzp,Takahashi:2017pje,Langlois:2018jdg} theories, which include Horndeski as a special case. 

It is well-known that, in general, there are many examples of static hairy black hole solutions to these theories. For example, the class of solutions admitting a hairy profile with a radially dependent scalar field $\phi(r)$ is well understood~\cite{Sotiriou:2013qea,Sotiriou:2014pfa,Babichev:2016rlq,Benkel:2016rlz,Babichev:2017guv,Lehebel:2017fag,Minamitsuji:2018vuw,BenAchour:2019fdf,Minamitsuji:2019tet}, and the general situation is analyzed in~\cite{Kobayashi:2012kh,Kobayashi:2014wsa}, with~\cite{Franciolini:2018uyq} providing an analysis from the effective field theory point of view. 

While these static solutions are interesting and yield powerful constraints, an equally important class of solutions are the cosmological ones, which are, of necessity, time-dependent. Among the allowed possibilities, some classes of shift-symmetric Horndeski theories admit particularly distinctive cosmological features, and are worthy of closer study. Recent studies of these theories have uncovered various hairy black hole solutions with non-trivial scalar profiles~\cite{Babichev:2013cya,Kobayashi:2014eva,Babichev:2016kdt,Babichev:2017lmw,BenAchour:2018dap,Motohashi:2019sen,Takahashi:2019oxz,Minamitsuji:2019shy,Minamitsuji:2019tet}, in which the scalar field depends linearly on time, as in the well-known example of the black hole solution~\cite{Mukohyama:2005rw} in ghost condensate theory~\cite{ArkaniHamed:2003uz}. Note that all of these hairy black hole solutions evade the no-hair theorem for Galileons proven in~\cite{Hui:2012qt}, since the time-dependent scalar field does not inherit all symmetries of the space-time. While these time-dependent solutions are intriguing, recent analyses have established that many of them suffer from fatal instabilities~\cite{Ogawa:2015pea,Babichev:2017lmw,Babichev:2018uiw,deRham:2019gha}, implying that they cannot be realized in the real world.

In this work we focus on a subclass of possible Horndeski theories of particular interest in cosmology --- those with a shift symmetry for the scalar field. We construct new time-dependent hairy black hole solutions to these theories, completing the taxonomy of such examples within the Horndeski class. We then carry out a comprehensive analysis of perturbations around these solutions, and demonstrate that they also exhibit an inevitable instability. Taken together with earlier results, we conclude that our calculations rule out the possibility of scalar hair with linear time dependence in shift-symmetric Horndeski theories.

\section{Black hole solutions in shift-symmetric Horndeski}

We consider the most general Horndeski theory~\cite{Horndeski:1974wa} with shift symmetry $\phi \to \phi + {\rm constant}$,
\be
S = \int \rd^4x \sqrt{-g} \Big[ {\cal L} _2 +{\cal L} _3+{\cal L} _4+{\cal L} _5 \Big]\,, 
\label{Horndeski}
\ee
with
\bea
{\cal L}_2 & = & P(X)\,; \nonumber \\
{\cal L}_3 & = & G_3(X) \square \phi\,; \nonumber \\
{\cal L} _4 & = &G_4(X) R + G_{4,X}(X)\Big[(\square \phi)^2 - \phi_{\mu\nu}\phi^{\mu\nu} \Big]\,; \nonumber \\
{\cal L}_5 & = & G_5(X)  G_{\mu\nu} \phi^{\mu\nu} - \frac{1}{6} G_{5,X}(X) \Big[ (\square \phi)^3 +2 \phi_{\mu}^{\;\; \nu} \phi_{\nu}^{\;\; \alpha} \phi_{\alpha}^{\;\; \mu} -3 \phi_{\mu\nu} \phi^{\mu\nu} \square \phi \Big]\,,
\label{L_i}
\eea
where $X\equiv -\frac{1}{2}g^{\mu\nu} \partial_{\mu} \phi \partial_{\nu}\phi$, and $\phi_{\mu\nu} \equiv \nabla_{\mu} \nabla_{\nu} \phi$.\footnote{
Conventions:  We work in mostly plus signature and use the curvature conventions $R^{\rho}{}_{\sigma\mu\nu}=\partial_{\mu}\Gamma^{\rho}_{\nu\sigma}+\ldots$ and $R_{\mu\nu}=R^{\rho}{}_{\mu\rho\nu}$. We denote the reduced Planck mass by $\Mpl=(8\pi G_{\rm N})^{-1/2}$.} The function $P(X)$ is chosen such that it
admits a ghost condensate background solution $X=\bar{X}={\rm constant}$~\cite{ArkaniHamed:2003uy}, for which
\begin{equation}
P(\bar{X}) =P_{,X}(\bar{X}) = 0 \,.
\end{equation}
As a specific example, one could choose $P(X) = \left(X - \bar{X}\right)^2$. With this choice, the theory containing~${\cal L}_2$ alone is already of cosmological interest, since inflation can be driven by the kinetic energy of the field~\cite{ArkaniHamed:2003uz} and gives rise to distinctive non gaussianities~\cite{ArkaniHamed:2003uz,Chen:2006nt}. In the late universe, perturbations around the ghost condensate in an isotropic, homogeneous cosmology then behave like dark matter~\cite{ArkaniHamed:2005gu}.

Most relevant to our analysis are the ghost condensate black hole solutions studied in~\cite{Mukohyama:2005rw}, where it was pointed out that perturbations around such black hole backgrounds are unstable, as they also are in flat space. More precisely, the kinetic matrix of perturbations is degenerate, and therefore the sound speed vanishes. Treated as an effective theory, it is then possible to include higher derivative terms that stabilize the dispersion relation.

In the broader context of GLPV~\cite{Gleyzes:2014dya} and DHOST~\cite{Langlois:2015cwa,Crisostomi:2016czh,BenAchour:2016fzp,Takahashi:2017pje,Langlois:2018jdg} theories, several further classes of hairy black hole solutions have been discovered~\cite{Babichev:2013cya,Kobayashi:2014eva,Babichev:2016kdt,Babichev:2017lmw,BenAchour:2018dap}. In the case of shift-symmetric quadratic DHOST theories, new classes of black hole solutions were derived and further investigated in~\cite{Motohashi:2019sen}, where it was shown that under the conditions 
\begin{equation}
G_{3,X}(\bar{X})  = G_{4,X}(\bar{X})  = G_{4,XX}(\bar{X})  = G_{5,X}(\bar{X})  = G_{5,XX}(\bar{X}) =0\,,
\end{equation}
the equation of motion for $\phi$ is automatically solved, while the Einstein equations reduce to 
\begin{equation}
G_{\mu\nu} = \frac{1}{2 G_4(\bar{X})}\Big(P(\bar{X}) g_{\mu\nu}  + P_{,X}(\bar{X})  \partial_{\mu} \phi \partial_{\nu}\phi \Big)\,.
\end{equation}
With $P_{,X}(\bar{X}) = 0$, these are immediately recognized as the Einstein equations with effective cosmological constant $\Lambda_{\rm eff} = -\frac{P(\bar{X})}{2 G_4(\bar{X})}$.
Therefore for a suitable choice of $\bar{P}$, dS/AdS Schwarzschild black holes are solutions to the theory~\eqref{Horndeski}. 

Alas, as shown in~\cite{deRham:2019gha}, such black hole solutions also suffer from the same problem that the kinetic matrix vanishes on the background, indicating
once again that the solution lies in the strongly coupled regime where the effective theory cannot be trusted. There is, however, a possible loophole. As already pointed
out in~\cite{Motohashi:2019sen}, the conditions $G_{4,X}(\bar{X})  = G_{4,XX}(\bar{X}) = 0$ are in fact unnecessary --- black hole solutions exist for an arbitrary choice of $G_4(X)$.
(Below we will show explicitly how this comes about.) 

Our goal in this paper is to capitalize on this additional freedom and to seek new black hole solutions with arbitrary~$G_4(X)$. We hope that in doing so the strongly
coupled regime can be avoided, and we find that having a general~$G_4(X)$ does indeed result in a non-degenerate kinetic matrix, as desired. However, unfortunately
one perturbation mode around these background solutions unavoidably propagates with imaginary sound speed, signaling a gradient instability. Although we will demonstrate this with a complete stability analysis, including both scalar and metric perturbations, the instability can already be seen in the decoupling limit where mixing with gravity is ignored, as discussed in Sec.~\ref{sec:bgdsclrpert}. 

\subsection{Black hole solutions with arbitrary $G_4(X)$}
\label{BH soln arbitrary G4}

The scalar and gravitational equations for the shift-symmetric Horndeski theory~\eqref{Horndeski} are respectively given by
\begin{align}
&{\cal E}^{\phi}_2+{\cal E}^{\phi}_3+{\cal E}^{\phi}_4+{\cal E}^{\phi}_5 =0\,;  \nonumber \\
&T_{2 \mu\nu}+T_{3 \mu\nu}+T_{4 \mu\nu}+T_{5 \mu\nu} =0\,,
\end{align}
where ${\cal E}^{\phi}_i$ and $T_{i \mu\nu}$ derive from the corresponding ${\cal L}_i$ in~\eqref{L_i}. The explicit expressions for~${\cal E}^{\phi}_3$,~${\cal E}^{\phi}_5$,~$T_{3 \mu\nu}$, 
and~$T_{5 \mu\nu}$, which derive from~${\cal L}_3$ and~${\cal L}_5$, will be of secondary interest to us and can be found in Appendix~\ref{sec:L3L5eom}.
Suffice to say that their vanishing requires 
\be
G_{3,X}(\bar{X})  = G_{5,X}(\bar{X})  = G_{5,XX}(\bar{X}) =0
\label{eqn:bkgcondition 3 and 5}
\ee
for some $\bar{X} = {\rm constant}$. As mentioned earlier, however, these conditions also imply that~${\cal L}_3$ and~${\cal L}_5$ do not contribute to the sound speed~\cite{deRham:2019gha}.
For this reason we will be primarily interested in~${\cal L}_2$ and~${\cal L}_4$. 

The contributions of~${\cal L}_2$ and~${\cal L}_4$ to the scalar equation of motion are given by 
\bea
\nonumber
{\cal E}^{\phi}_2 &=& \nabla_{\mu}\left( P_{,X} \partial^{\mu}\phi\right)\,;  \\
{\cal E}^{\phi}_4 &=& \nabla_{\mu}\Big( \Big( G_{4,X}R + G_{4,XX}\left[(\square \phi)^2 - \phi_{\mu\nu}\phi^{\mu\nu} \right] \Big) \partial^{\mu}\phi + 2\nabla_{\nu}\big( G_{4,X}\left[ \square \phi g^{\mu\nu} - \phi^{\mu\nu} \right] \big) \Big)\,,
\label{E2E4}
\eea
while their contributions to the energy-momentum tensor are
\bea
\nonumber
T_{2\mu\nu} &=& g_{\mu\nu} P + P_{,X}  \partial_{\mu} \phi \partial_{\nu}\phi  \,; \\ 
\nonumber
T_{4\mu\nu} &=& g_{\mu\nu} \Big[  G_4 R  + G_{4,X}\big(\phi_{\rho\sigma}\phi^{\rho\sigma} - (\square \phi)^2 + 2 R_{\rho\sigma}  \partial^{\rho} \phi \partial^{\sigma}\phi \big )   -2 G_{4,XX} \big( \partial_{\rho} X\partial^{\rho} X +\square \phi \partial_{\rho}\phi \partial^{\rho} X \big) \Big] \\
\nonumber
& & -~2 G_4 R_{\mu\nu}   -2G_{4,X} \Big( \phi^{\rho}{}_{ \mu} \phi_{\nu\rho} + 2 \partial_{(\mu} \phi R_{\nu)\rho} \partial^{\rho} \phi + R_{\rho\mu \sigma \nu} \partial^{\rho}\phi \partial^{\sigma}\phi \Big) \\
\nonumber 
&  & +~2 \Big(G_{4,X} \square \phi + G_{4,XX} \partial_{\rho}\phi \partial^{\rho} X  \Big) \phi_{\mu\nu} + \Big( G_{4,X} R + G_{4,XX} \big( (\square \phi)^2-\phi_{\rho\sigma}\phi^{\rho\sigma} \big) \Big)\partial_{\mu} \phi \partial_{\nu}\phi  \\
& &+~2 G_{4,XX} \Big(  2\square \phi \partial_{(\mu} \phi \partial_{\nu)} X  +  \partial_{\mu} X \partial_{\nu} X  - 2 \partial^{\rho} X \phi_{\rho (\mu } \partial_{\nu)} \phi \Big) \,.
\label{T2T4}
\eea  
It is easy to see that ${\cal E}^{\phi}_2$ vanishes on a black hole background with $X = \bar{X} = {\rm constant} $ provided that $P_{,X}(\bar{X}) =0$,
while $T_{2\mu\nu}$ vanishes if, furthermore, $P(\bar{X}) =0$. Henceforth let us denote the constant $X$ solution as
\be
\bar{X} \equiv \frac{1}{2}\Lambda^4\,.
\label{barX Lambda}
\ee

Following~\cite{Mukohyama:2005rw} it is convenient to work in Lema\^{i}tre coordinates, since these greatly simplify the analysis compared to Schwarzschild coordinates. For a static Schwarzschild black hole these take the form
\begin{equation} \label{eqn:lemaitreg}
  \rd s^2 = -\rd\tau^2 + \frac{r_{\rm s}}{r} \rd \rho^2 + r^2 \rd\Omega_2^2\,,
  \end{equation}
where $r$ is the usual Schwarzschild radial coordinate, and
\be
\tau = t + 2\sqrt{r r_{\rm s}} + r_{\rm s} \ln\left| \frac{\sqrt{r} - \sqrt{r_{\rm s}} }{\sqrt{r} + \sqrt{r_{\rm s}}}\right| \,;\qquad \rho = \tau + \frac{2}{3} r_{\rm s} \left(\frac{r}{r_{\rm s}}\right)^{3/2}\,.
\label{taurho}
\ee
These imply
\begin{equation}  
  \qquad  r= \left[ \frac{3}{2}(\rho-\tau)\right]^{2/3} r_{\rm s}^{1/3}\,.
\end{equation}
(In Appendix~\ref{sec:AppdxA} we give the general form of a static, spherically-symmetric space-time in Lema\^{i}tre-type coordinates.)
The advantage of Lema\^{i}tre coordinates is that~\eqref{barX Lambda} can be satisfied simply by a linear time-dependent profile:
\begin{equation}
 \bar{\phi}(\tau) =  \Lambda^2 \tau \,.
 \label{linearphi}
\end{equation}

It remains to show that the ${\cal L}_4$ contributions to the equations of motion vanish on this background.
To do so, we will make use of the following two identities satisfied by the linear profile~\eqref{linearphi}:
\begin{subequations}
\label{wholeset}
\bea
\label{identity1}
& \displaystyle   \nabla_{\nu} \left( \square \bar{\phi} g^{\mu\nu} - \bar{\phi}^{\mu\nu}  \right) = 0\,; & \\
\label{identity2}
& \square \bar{\phi} \bar{\phi}_{\mu\nu}  -  \bar{\phi}^{\rho}{}_{ \mu} \bar{\phi}_{\nu\rho}  - R^{\rm BH}_{\rho\mu \sigma \nu} \partial^{\rho}\bar{\phi} \partial^{\sigma}\bar{\phi}  = 0\,,
\eea
\end{subequations}
where $R^{\rm BH}_{\rho\mu \sigma \nu} $ is the Riemann tensor for the metric~\eqref{eqn:lemaitreg}. The trace of the latter gives
\be
(\square \bar{\phi} )^2-\bar{\phi}_{\rho\sigma} \bar{\phi}^{\rho\sigma} =0 \,.
\label{identity2trace}
\ee
Using~\eqref{wholeset}$-$\eqref{identity2trace}, together with $X = {\rm constant}$ and $R^{\rm BH} = 0$, it is easy to see that~${\cal E}^{\phi}_4$ and~$T_{4\mu\nu}$ vanish. 
Interestingly, nothing in this analysis so far constrains the function $G_4(X)$ or its derivatives --- a Schwarzschild black hole with linear profile~\eqref{linearphi} is an exact
solution to the field equations {\it for arbitrary $G_4(X)$}.\footnote{If the Horndeski theory is regarded as an effective field theory, the black hole entropy can provide a constraint on effective operators. The Wald entropy~\cite{Wald:1993nt} of black holes in any covariant modified gravity theory is given by 
\begin{equation}
   S_{\rm Wald} = \oint_{\Sigma} \frac{\delta L}{ \delta R^{\mu\nu\rho \sigma}} \epsilon^{\mu\nu} \epsilon^{\rho\sigma}  ,\quad    \nabla_{\mu}\xi_{\nu} \big|_{\Sigma} = \kappa \epsilon_{\mu\nu},
\end{equation}
where $\Sigma $ is the bifurcation surface and $\kappa$ is the surface gravity. Evaluated on the above solution,
\begin{equation}
 S =\frac{A}{4G_N}\left( \frac{ 2 }{\Mpl^2} \bar{G}_4(\bar{X} ) \right)  ,
\end{equation}
where $A=4\pi r_{\rm s}^2$ is the area of the horizon. It is argued that in a healthy effective field theory of gravity, the correction to the Wald entropy must be positive compared to pure GR~\cite{Cheung:2018cwt}. The spirit behind this is closely related to unitarity and causality in the UV complete theory \cite{Hamada:2018dde}. Comparing with $G_4 =  \frac{\Mpl}{2}$ in GR, this demands that $\bar{G}_4 \geq \frac{\Mpl^2}{2}$. }  

Incidentally, there exist other solutions to the field equations describing a Schwarzschild black hole~\eqref{eqn:lemaitreg} and scalar field profile satisfying $X={\rm constant}$.  
For instance, instead of~\eqref{linearphi} another possibility is 
\be
\phi(r)= \Lambda^2\left[ \sqrt{r(r-r_{\rm s})} + r_{\rm s} {\rm tanh}^{-1}\left(1 -\frac{r_{\rm s}}{r}\right) \right]\,.
\label{standard profile}
\ee
However, this solution does not satisfy the identities~\eqref{wholeset}$-$\eqref{identity2trace}. Requiring this type of ansatz to solve the Einstein equations further requires $G_{4,X}(\bar{X}) = G_{4,XX}(\bar{X})=0$, which, as mentioned earlier, results in a vanishing sound speed. 

\subsection{Scalar perturbations around fixed black hole background} \label{sec:bgdsclrpert}

Before carrying out a complete stability analysis, it is instructive to examine the stability of the scalar profile~\eqref{linearphi} on a fixed Schwarzschild black hole background. In other words,
we ignore the backreaction of scalar perturbations. One reason to do this is that the analysis is significantly simpler than the full gravitational treatment. Another reason is that the approximation of ignoring dynamical gravity is reasonable in decoupling regions of parameter space where gravitational mixing can be neglected, as detailed below, as well as in the case of $\ell =1$ metric perturbations, where tensorial perturbations are expected to be non-dynamical.

Perturbing~\eqref{linearphi} around the fixed Schwarzschild background,
\be
  \phi(\tau,\vec{x}) = \Lambda^2\Big(\tau + \pi(\tau,\vec{x})\Big)\,,
\ee
the action for the perturbation $\pi(\tau,\vec{x})$ at quadratic order is
\be
{\cal L}^\pi_{\rm quad}  = \Lambda^4\left(\bar{P}_{,XX} \bar{g}^{\tau\tau} (\partial_{\tau} \pi)^2  + 4\bar{G}_{4,XX} \frac{r_{\rm s}}{r^3}  \bar{g}^{\rho \rho}(\partial_{\rho} \pi)^2  -2\bar{G}_{4,XX} \frac{r_{\rm s}}{r^3}  \bar{g}^{AB}_{\Omega_2} \partial_A \pi \partial_B \pi\right)\,,
\ee
where $\bar{P}_{,XX} \equiv P_{,XX}(\bar{X})$ {\it etc.}, and the inverse metric components can be read off from~\eqref{eqn:lemaitreg}. In particular, $\bar{g}^{AB}_{\Omega_2}$ is the metric on the two-sphere, with $A, B$ indices denoting the angular variables~$\theta,\varphi$. The radial and angular sound speeds can be immediately identified:
\be
 c_{\rho}^2 = 4\frac{\bar{G}_{4,XX}}{P_{,XX}} \frac{r_{\rm s}}{r^3} \,; \qquad  c_{\theta,\varphi}^2 = -2\frac{\bar{G}_{4,XX}}{P_{,XX}} \frac{r_{\rm s}}{r^3}\,.
\label{cs decoupling}
\ee
Note that the sound speeds vanish for $\bar{G}_{4,XX} = 0$, as must be the case for the standard solution~\eqref{standard profile}.

As hoped, allowing for general $G_4(X)$ results in non-vanishing propagation speeds. Unfortunately, as is clear from~\eqref{cs decoupling}, the relative sign between
radial and angular squared sound speeds is negative, 
\be
c_{\rho}^2 = -2 c_{\theta,\varphi}^2\,,
\label{crho angular prop}
\ee
indicating an instability irrespective of the choice of~$G_4(X)$. This is suggestive, but not conclusive, since these expressions will inevitably be altered when taking into account perturbations of the metric tensor. Nevertheless we anticipate~\eqref{cs decoupling} to hold approximately in parametric regimes where gravitational mixing can be ignored. 

Indeed, the comprehensive analysis using scalar-vector-tensor (SVT) decomposition on the two-sphere will be done in the next Section. Looking ahead at the exact result~\eqref{cs angular}, with $g_1$ and $g_2$ defined in~\eqref{g1def} and~\eqref{g2def}, respectively, we see that the scalar sound speeds reduce to~\eqref{cs decoupling} in the limit 
\be
\bar{G}_4 \gg \Lambda^4 \bar{G}_{4,X}\,,\;\Lambda^8\bar{G}_{4,XX}\,,\; \frac{\bar{G}_{4,X}^2}{\bar{G}_{4,XX}}\,.
\label{decoupling limit}
\ee
This intuitively makes sense --- from~\eqref{L_i} one can interpret $\bar{G}_4$ as setting the effective Planck scale, and the above states that a large $\bar{G}_4$ ensures decoupling. 
Interestingly, while $c_{\rho}^2$ and $c_{\theta,\varphi}^2$ individually receive gravitational corrections in the complete analysis, we will find that the relation~\eqref{crho angular prop}
is preserved in the $\ell=1$ sector, with the same proportionality constant. Intuitively, this is because tensorial metric perturbations are non-dynamical in the~$\ell=1$ sector. 
 
\section{Full metric perturbation analysis: Parity-odd sector}

We now carry out the full perturbation analysis, taking into account both perturbations of the scalar field, and those of the metric, to which they are unavoidably coupled. 
The perturbed metric is 
\begin{equation}
g_{\mu\nu} = \bar{g}_{\mu\nu} +h_{\mu\nu} \,,
\end{equation}
where $\bar{g}_{\mu\nu}$ is the background black hole metric~\eqref{eqn:lemaitreg}, and $h_{\mu\nu}$ is the perturbation. The perturbed scalar field is
\be
\phi(\tau,\vec{x}) = \Lambda^2 \big(\tau + \pi(\tau,\vec{x})\big)\,.
\label{scalar pert def}
\ee

Since our background enjoys spherical symmetry,~$h_{\mu\nu}$ can be decomposed into scalar, vector and tensor harmonics on the two-sphere~\cite{Regge:1957td}. Just like the scalar-vector-tensor (SVT) decomposition in~$\mathbb{R}^3$ can be conveniently done in terms of Fourier components, the SVT decomposition on the two-sphere is carried out using spherical harmonics~$Y_\ell^m(\theta,\varphi)$. As reviewed in Appendix~\ref{sec:appdxB}, vector and tensor spherical harmonics can be expressed in terms of derivatives of~$Y_\ell^m$. For example $h_{\tau B}$ is a vector on the two-sphere such that it can be decomposed as $h_{\tau A} = \sum_{\ell,m}\left[ \alpha^{\ell m} \nabla_{A}+ h_0^{\ell m} \epsilon_A^{\;\;\; B} \nabla_B  \right]Y^m_{\ell} $, where $\epsilon_{AB}$ and $\nabla_A$ are the Levi-Civita tensor and covariant derivative on the two-sphere, respectively. 

Under the parity transformation $(\theta,\varphi) \to (\pi-\theta,\varphi+\pi)$, each term in the SVT expansion for~$h_{\mu\nu}$ picks up either a factor of $(-1)^\ell$ or $(-1)^{\ell+1}$. 
Hence these are referred to respectively as even and odd perturbations~\cite{Regge:1957td}. Following this convention, the metric perturbation can be decomposed into
even-sector and odd-sector pieces:
\begin{equation}
h_{\mu\nu} =  h^{\rm odd}_{\mu\nu} +h^{\rm even}_{\mu\nu}\,.
\label{h even odd}
\end{equation}
Because the theory~\eqref{Horndeski} is parity invariant, these decouple at linear order. Meanwhile, the scalar perturbation~$\pi$ belongs to the parity-even sector. 

As usual, $h_{\mu\nu}$ is not unique, as it changes under small diffeomorphisms. Analogously to~\eqref{h even odd}, a general diffeomorphism vector $\xi^{\mu}$ can also be decomposed into odd-sector and even-sector parts, 
\be
\xi^{\mu} = \xi^{\mu}_{\rm odd} + \xi^{\mu}_{\rm even}\,.
\label{diff even odd}
\ee
In this Section we begin with the parity-odd sector, while the analysis of the even sector will be discussed in Sec.~\ref{parity even sec}.

We are fortunate that the perturbation analysis for the odd sector has already been carried out~\cite{Ogawa:2015pea,Takahashi:2019oxz}. Nevertheless, it is instructive to repeat the analysis here, using instead our preferred Lema\^{i}tre coordinates. We will see that working with Lema\^{i}tre coordinates simplifies things significantly, and it is this simplification that will enable us to perform the even sector analysis below. 

In this basis, a general odd-sector metric perturbation takes the form
\begin{equation} \label{eqn:oddpert}
  h^{\rm odd}_{\mu\nu} = \sum_{\ell,m} \begin{pmatrix}
  0& 0& h^{\ell m}_0 \epsilon_A^{\;\;\; B} \nabla_B  \\
  0& 0& h^{\ell m}_1 \epsilon_A^{\;\;\; B} \nabla_B \\
  h^{\ell m}_0 \epsilon_A^{\;\;\;B} \nabla_B&  h^{\ell m}_1 \epsilon_A^{\;\;\;B} \nabla_B & h^{\ell m}_2 \epsilon_{(A}^{\quad C} \nabla_{B)} \nabla_C 
  \end{pmatrix} Y_{\ell}^m(\theta , \varphi)\,.
\end{equation}
Each term in the sum has parity $(-1)^{\ell+1}$ under $(\theta,\varphi) \to (\pi-\theta,\varphi+\pi)$. Meanwhile, the parity-odd part of the diffeomorphism vector~\eqref{diff even odd} can be expressed as 
\begin{equation}
\xi^{\mu}_{\rm odd} =\sum_{\ell, m} \left(0,0,\xi^{\ell m} \epsilon^{AB}\nabla_{B}\right)Y_{\ell}^m(\theta , \varphi) \,.
\end{equation}
Under such a diffeomorphism, $x^{\mu} \to x^{\mu} + \xi^{\mu}_{\rm odd}$, the metric functions~$h^{\ell m}_0$,~$h^{\ell m}_1$ and~$h^{\ell m}_2$ transform as
\bea
\nonumber
\delta h_0 &=& -\dot{\xi} +2\frac{\dot{r}}{r} \xi\,;  \\
\nonumber
\delta h_1 &=& -\xi' +2 \frac{r'}{r} \xi\,; \\
\delta h_2 &=& -2 \xi \,,
\label{h_1 transform}
\eea
where we have suppressed $\ell$, $m$ indices for simplicity. Furthermore, here and henceforth,~$(\;)'$ and~$(\;)\dot{}$ denote $\rho$ and~$\tau$ differentiation, respectively.
We use this gauge freedom to set 
\be
h_2 = 0\, ,
\ee
which is the Regge-Wheeler gauge. Furthermore, from~\eqref{h_1 transform} we can identify a gauge-invariant linear combination of~$h_0$ and~$h_1$:
\be
h_0' -\dot{h}_1 - 2 \frac{r'}{r} h_0+ 2 \frac{\dot{r}}{r} h_1 = h_0' -\dot{h}_1 -2 \sqrt{\frac{r_{\rm s}}{r^3}} (h_0 + h_1) \,,
\label{gauge inv comb}
\ee
where in the last step we have used~\eqref{taurho}. Note that this gauge-invariant quantity is identical to that of pure gravity~\cite{Martel:2005ir},
since scalar perturbations do not enter the calculation. 

After integrating out the angular coordinates, the quadratic action for odd perturbations reduces to 
\be
{\cal L}_{\rm odd} = \frac{g_1}{2}  \sqrt{\frac{r}{r_{\rm s}}} \left(h_0' -\dot{h}_1 -2 \sqrt{\frac{r_{\rm s}}{r^3}} (h_0 + h_1) \right)^2 
+\frac{\ell(\ell+1) -2}{2r^2}\left( g_1\sqrt{\frac{r_{\rm s}}{r}}  h_0^2 - \bar{G}_4\sqrt{\frac{r}{r_{\rm s}}} h_1^2\right)  \,,
\label{LoddwithoutPsi}
\ee
where the implicit $(\ell,m)$ indices are summed over, and where we have defined
\be
g_1 \equiv \bar{G}_4 - \Lambda^4 \bar{G}_{4,X}\,.
\label{g1def}
\ee
The quantity inside the parentheses in the first term is immediately recognized as the gauge-invariant combination~\eqref{gauge inv comb}. We exploit this by 
introducing an auxiliary field $\Psi$ into the Lagrangian,
\be
{\cal L}_{\rm odd} = g_1 \sqrt{\frac{r}{r_{\rm s}}}  \bigg[\Psi \left(h_0' -\dot{h}_1 -2 \sqrt{\frac{r_{\rm s}}{r^3}} (h_0 + h_1)  \right)  -\frac{\Psi^2}{2} \bigg]
+\frac{\ell(\ell+1) -2}{2r^2}\left( g_1\sqrt{\frac{r_{\rm s}}{r}}  h_0^2 - \bar{G}_4\sqrt{\frac{r}{r_{\rm s}}} h_1^2\right)\,.
\label{LoddwithPsi}
\ee
The equation of motion for $\Psi$ sets it equal to the gauge-invariant combination~\eqref{gauge inv comb}, and substitution back into~\eqref{LoddwithPsi} reproduces the original action~\eqref{LoddwithoutPsi}. Instead, we express $h_0$ and $h_1$ in terms of $\Psi$ through their equations of motion,
\bea
\nonumber
 h_0 &=& \frac{1}{\ell(\ell+1) -2} \,\frac{r^2}{r_{\rm s}} \left(r \Psi' + \frac{5}{2} \sqrt{\frac{r_{\rm s}}{r}}\Psi \right)\,; \\
 h_1 &=&  \frac{1}{\ell(\ell+1) -2} \frac{g_1}{\bar{G}_4} r\left(r \dot{\Psi} - \frac{5}{2} \sqrt{\frac{r_{\rm s}}{r}} \Psi \right)\,.
\label{h0 h1 Psi}
\eea
This is valid for $\ell \geq 2$. The case $\ell =1$ must be treated separately, since the dipole components of the two degrees of freedom $h_0$ and $h_1$ correspond 
to a pure gauge mode and a small rotation into Kerr black hole~\cite{Martel:2005ir}. Substituting~\eqref{h0 h1 Psi} back into the action~\eqref{LoddwithPsi}, we obtain
\be
{\cal L}_{\rm odd} =   \frac{g_1}{2} \sqrt{\frac{r}{r_{\rm s}}} \frac{1}{\ell(\ell+1)-2} 
\Bigg\{r^2 \left(\frac{g_1}{\bar{G}_4} \dot{\Psi}^2 - \frac{r}{r_{\rm s}} \Psi'^2\right)  + \frac{1}{4} \left[15\frac{g_1}{\bar{G}_4} \frac{r_{\rm s}}{r} + 3 - 4\ell (\ell + 1) \right] \Psi^2 \Bigg\}   \,.
\label{Lodd final}
\ee 

Thus the parity-odd perturbations are encoded in the gauge-invariant mode functions~$\Psi$.\footnote{In the special case $G_{4,X}=0$ ({\it i.e.}, $g_1=G_4$), the
action~\eqref{Lodd final} reduces to the GR result, and its equation of motion corresponds to a Regge-Wheeler type equation in Lema\^{i}tre coordinates.} 
Inspection of~\eqref{Lodd final} allows us to determine the stability of this sector. The coefficient of $\dot{\Psi}^2$ is manifestly positive, hence there is no ghost. 
From the form of the kinetic term, and recalling that $g^{\rho\rho} = \frac{r}{r_{\rm s}}$, we can read off the radial sound speed 
\be
c_\rho^2 = \frac{\bar{G}_4}{g_1} = \frac{\bar{G}_4}{\bar{G}_4 - \Lambda^4 \bar{G}_{4,X}}\,.
\ee
Requiring subluminal propagation imposes the constraint
\be
\Lambda^4 \bar{G}_{4,X}  < 0\,.
\label{crho bound odd sector}
\ee
The third stability requirement is that the ``mass" term in~\eqref{Lodd final} has the correct sign, but this is guaranteed for all $\ell \geq 1$ once~\eqref{crho bound odd sector} is satisfied. 
Additionally, there are of course tight observational limits on the propagation speed of gravitational waves from neutron star mergers~\cite{TheLIGOScientific:2017qsa,GBM:2017lvd,Monitor:2017mdv}.

\section{Parity-even sector}
\label{parity even sec}

We now turn to even-sector perturbations. Working in Lema\^{i}tre coordinates is critical here --- the calculation below, while still formidable in Lema\^{i}tre coordinates, 
would be nearly impossible in Schwarzschild coordinates. 

Analogously to~\eqref{eqn:oddpert}, a general parity-even metric perturbation $h^{\rm even}_{\mu\nu}$ can be parameterized as
\be  
\label{eqn:gevenpert}
 h^{\rm even}_{\mu\nu} = \sum_{\ell, m} \begin{pmatrix}
 H_0^{\ell m} & ~~~~H_1^{\ell m} &  ~~~~\alpha^{\ell m} \nabla_A  \\
  H_1^{\ell m} & ~~~~\frac{r_{\rm s}}{r} H_2^{\ell m}&  ~~~~\beta^{\ell m}  \nabla_A  \\
  \alpha^{\ell m}  \nabla_A & ~~~~\beta^{\ell m}  \nabla_A & ~~~~r^2K^{\ell m}g_{AB}  + Q ^{\ell m} \nabla_A \nabla_B    \\
 \end{pmatrix} Y_{\ell}^m(\theta , \varphi)\,, \\
\ee
where $g_{AB}$ is the metric tensor on the two-sphere. The metric coefficients are all scalar functions of the Lema\^{i}tre coordinates~$\rho$ and~$\tau$. 
Each term has parity $(-1)^{\ell}$ under $(\theta,\varphi) \to (\pi-\theta,\varphi+\pi)$. Similarly, the scalar perturbation~$\pi$ defined in~\eqref{scalar pert def}
can be expanded as
\be
\pi = \sum_{\ell , m} \pi^{\ell m}(\tau ,\rho) Y_{\ell}^m(\theta,\varphi)\,,
 \ee
with each term having parity~$(-1)^\ell$. 

The parity-even part of the diffeomorphism vector~\eqref{diff even odd} can be expressed as 
\be
 \xi^{\mu}_{\rm even}  =\sum_{\ell,m} \left( {\cal T}^{\ell m}, \; {\cal R}^{\ell m}, \;\Theta^{\ell m} g^{AB}\nabla_B \right) Y_{\ell}^m(\theta,\varphi)\,,
\ee
where ${\cal T}^{\ell m}$, ${\cal R}^{\ell m}$ and $\Theta^{\ell m}$ are all functions of $(\tau, \rho)$. Under a such diffeomorphism, $x^{\mu} \to x^{\mu} + \xi^{\mu}_{\rm even}$, the metric coefficients transform as 
\begin{gather}
\begin{aligned}
\label{eqn:evendiff}
\delta H_0 &= 2\dot{\cal T} \,; \qquad\qquad &\delta H_1&= {\cal T}' - \frac{r_{\rm s}}{r} \dot{\cal R}\,; \qquad\qquad \delta H_2 = \frac{ \dot{r}}{r} {\cal T} + \frac{ r' }{r} {\cal R} - 2 {\cal R'}\,; \\
\delta \alpha & = {\cal T} - r^2 \dot{\Theta}\,; \qquad\qquad &\delta \beta &= - \frac{r_{\rm s}}{r} {\cal R} - r^2 \Theta '\,;  \\
\delta K &=  -2 \frac{\dot{r}}{r}{\cal T} - 2 \frac{r'}{r} {\cal R}  \,;\qquad \qquad &\delta Q &= -2 \Theta\,,
\end{aligned}
\end{gather}
where $\ell$, $m$ indices have been suppressed to avoid clutter. As in the odd sector, it is once again possible to identify gauge-invariant combinations.
However we will find it more convenient to use the above gauge freedom to pick a suitable gauge, making sure that no constraint is lost in the process.

In what follows we will treat the monopole~($\ell=0$) and dipole~($\ell=1$) cases separately, followed by a general analysis for arbitrary~$\ell$. Because the
constraints are significantly more complex in the general case, we will only be able to provide an incomplete analysis, which we hope can form the
basis of future investigations.

\subsection{Monopole perturbation}

The monopole ($\ell=0$) perturbation is a special case, as the variables $\alpha$, $\beta$ and $Q$ do not exist in this case. The gauge function $\Theta$ is also absent. The remaining two gauge functions ${\cal T}$ and ${\cal R}$ can be used to impose the gauge choice
\be
H_0=K= 0\,,
\ee
leaving us with $H_2$ and $\pi$ as degrees of freedom. The Lagrangian density then reduces to
\bea
{\cal L}_{\rm even}^{(\ell = 0)} &=&  \frac{1}{2} \bar{P}_{,XX} \Lambda^8 \sqrt{r_{\rm s} r^3} \dot{\pi}^2  + 2 \left(\bar{G}_4-2g_1 -g_2\right)\sqrt{\frac{r_{\rm s}}{r}} \delp'^2  \nonumber \\
&+& \frac{\bar{G}_4}{2} \sqrt{\frac{r_{\rm s}}{r}}H_2^2 + H_2\left( 2 (g_1 +g_2)r_{\rm s} \ddot{\pi} - 2(\bar{G}_4-g_1)\sqrt{\frac{r_{\rm s}}{r}} \delp' \right)  \nonumber\\
 &+& rH_1 \left(2 g_1  \dot{H}_2 -4(g_1+g_2) \dot{\pi}' \right)\,,
\label{Levenl=0}
\eea
where the implicit $(\ell,m)$ indices are summed over, and where we have defined
\be
g_2 \equiv -\bar{G}_4 + 2\Lambda^4 \bar{G}_{4,X}+ \Lambda^8 \bar{G}_{4,XX} \,.
\label{g2def}
\ee
Clearly, $H_1$ is a Lagrange multiplier, and the corresponding constraint can be solved by 
\begin{equation}
 H_2 = 2 \frac{g_1 +g_2}{g_1} \delp'\,. 
\label{H2 soln}
\end{equation}
Note that we have chosen the physical boundary condition such that there is no constant piece in time. (In fact, a constant shift in $H_2$ corresponds to a shift of integration constant of the background \cite{Kobayashi:2014wsa} which can be absorbed into $r_s$. We are not interested in a mere change of the background solution.) 

Substituting~\eqref{H2 soln} into the Lagrangian yields
\be
{\cal L}_{\rm even}^{(\ell = 0)} = \sqrt{\frac{r_{\rm s}}{r}} \left(\frac{1}{2} \bar{P}_{,XX} \Lambda^8 r^2 \dot{\pi}^2 + 2\frac{g_2}{g_1^2}\left(g_1^2 +\bar{G}_4 g_2\right) \delp'^2\right)\,.
\ee
From this we can infer that absence of ghost requires
\be
\bar{P}_{,XX} > 0\,,
\label{ghost free}
\ee
just as in the case of a ghost condensate, while gradient stability requires
\be
g_2\left(g_1^2 +\bar{G}_4 g_2\right) < 0\,.
\label{radial stability}
\ee
Using the fact that $g^{\rho\rho} = r_{\rm s}/r$, we obtain the radial sound speed\footnote{Note that the above sound speed vanishes in the special case $\bar{G}_{4,X}=\bar{G}_{4,XX}=0$, consistent with earlier results~\cite{deRham:2019gha}. In this limit the effective theory describing perturbations breaks down, as it exhibits a strong coupling problem.} 
\begin{equation}
  c_{\rho}^2 = -4 \frac{g_2(g_1^2 +\bar{G}_4g_2) }{g_1^2  \Lambda^8 \bar{P}_{,XX}}\frac{r_{\rm s}}{r^3} = 4 \frac{(\bar{G}_{4,X}^2 + \bar{G}_4 \bar{G}_{4,XX})(\bar{G}_4-2\Lambda^4 \bar{G}_{4,X} - \Lambda^8 \bar{G}_{4,XX})}{(\bar{G}_4-\Lambda^4 \bar{G}_{4,X})^2\bar{P}_{,XX}}\frac{r_{\rm s}}{r^3} \,.
\label{csho ell = 0}
\end{equation}
Below we will find an identical radial sound speed for $\ell = 1$, which makes sense, since scalar modes should propagate radially at the same speed independent of the multipole moment.
From~\eqref{csho ell = 0} we can derive a constraint by demanding subluminality. However, we will not bother to do so, because in the $\ell = 1$ case below we will discover a more worrisome pathology,
namely that it is impossible to satisfy $c_{\rho}^2 > 0$ while at the same time having stable propagation in the angular directions. 

\subsection{Dipole perturbation: instability of the scalar mode}

This subsection describes the main result of our paper. The dipole case, $\ell=1$, is a special case for a different reason. Although not {\it a priori} obvious, it turns out that the angular part of the metric perturbations, $h_{AB}$, is diagonal and depends only on the combination $r^2K-Q$~\cite{Martel:2005ir}. This allows us to choose a gauge in which 
\be
H_0=\beta=r^2K-Q=0\,,
\ee
leaving us with four component fields: $H_1$, $H_2$, $\alpha$ and $\pi$. 

Up to total derivatives, the resulting quadratic Lagrangian takes the form
\bea
{\cal L}^{(\ell=1)}_{\rm even} &=&  \frac{\Lambda^8}{2} \bar{P}_{,XX}  \sqrt{r_{\rm s} r^3} \dot{\pi}^2  + 2\big(\bar{G}_4-2g_1-g_2\big) \left(\frac{r_{\rm s}}{r}\right)^{3/2} \left(\frac{r}{r_{\rm s}}\delp'^2  - \frac{1}{r^2} \delp^2\right) - 2(g_1 +g_2 ) \frac{r_{\rm s}}{r^2} \alpha \dot{\pi}    \nonumber \\
& + &\frac{\bar{G}_4}{2}  \sqrt{\frac{r_{\rm s}}{r}} H_2 ^2+ H_2\left( 2 (g_1+g_2) r_{\rm s} \ddot{\delp} - 2(\bar{G}_4-g_1)\sqrt{r_{\rm s} r}\left( \frac{\delp}{r} \right)'  +2 g_1 \sqrt{\frac{r_{\rm s}}{r^3}}\left(r \alpha \right)^{\cdot} \right) \nonumber \\ 
&+& g_1 \sqrt{\frac{r}{r_{\rm s}}}H_1^2 + H_1\left( 2g_1 r \dot{H}_2 -4(g_1+g_2) r \dot{\pi}' - 2 g_1 \sqrt{\frac{r}{r_{\rm s}}} \alpha'\right) +  g_1 \sqrt{\frac{r}{r_{\rm s}}} \alpha'^2 \,,
\label{Levenl=1}
\eea
with implicit $(\ell,m)$ indices summed over. As in the monopole case, $H_1$ is once again non-dynamical (though not a Lagrange multiplier). Its equation of motion gives
\be
H_1 = \sqrt{\frac{r_{\rm s}}{r}} \left(-r \dot{H}_2 + 2 \frac{g_1 + g_2}{g_1} r\dot{\pi}' \right) + \alpha'\,.
\ee
Substituting this back into~\eqref{Levenl=1} and integrating by parts, it is easy to see that $\alpha$ becomes a Lagrange multiplier, with the constraint it imposes given by
\begin{equation}
 r \dot{H}_2' + 2 \sqrt{\frac{r_{\rm s}}{r}} \dot{H}_2+ \frac{3}{2}  \frac{ r_{\rm s}}{r^2} H_2  = \frac{g_1 + g_2}{g_1}\left(2\left( r\dot{\pi}'   \right)' -  \frac{r_{\rm s} }{r^2} \dot{\pi}\right)\,.
\label{alpha cons 1}
\end{equation}
Remarkably, by performing the field redefinition
\begin{equation}
h_2  \equiv H_2 - 2 \frac{g_1+g_2}{g_1} r \left( \frac{\delp}{r} \right)'\,,
\end{equation}
the constraint~\eqref{alpha cons 1} reduces to an equation for $h_2$ only:
\be
r \dot{h}_2' + 2 \sqrt{\frac{r_{\rm s}}{r}} \dot{h}_2+ \frac{3}{2}  \frac{ r_{\rm s}}{r^2} h_2   = 0\,.
\label{alpha cons}
\ee
 
Writing the Lagrangian in terms of $\delp$, $h_2$ and $\alpha$, we then obtain
\bea
 {\cal L}^{(\ell=1)}_{\rm even} &=& \frac{\Lambda^8}{2} \bar{P}_{,XX}  \sqrt{r_{\rm s} r^3} \dot{\pi}^2 + 2 \frac{g_2}{g_1^2} \left(g_1^2 +\bar{G}_4g_2\right) \left(\frac{r_{\rm s}}{r}\right)^{3/2} \left(\frac{r}{r_{\rm s}}\delp'^2  - \frac{1}{r^2} \delp^2\right) \nonumber \\
 &+& 2 (g_1 +g_2) \left( r_{\rm s} \dot{\pi}  + 3 \left(\frac{r_{\rm s}}{r}\right)^{3/2} \delp \right) \dot{h}_2 -   \frac{g_1^2 + \bar{G}_4 g_2}{g_1} \sqrt{\frac{r_{\rm s}}{r}} \delp   \left( 2 h_2'  + \sqrt{\frac{r_{\rm s}}{r^3}} h_2\right) \nonumber \\
 &-& g_1 \sqrt{r_{\rm s} r^3 } \dot{h}_2^2 + \frac{\bar{G}_4}{2}  \sqrt{\frac{r_{\rm s}}{r}} h_2^2  -2 g_1  \alpha\left(  r \dot{h}_2' + 2 \sqrt{\frac{r_{\rm s}}{r}} \dot{h}_2 + \frac{3}{2}  \frac{r_{\rm s}}{r^2} h_2\right)\,.
\label{L dipole pi h2}
\eea
Thus, the constraint~\eqref{alpha cons} imposes a particular form for $h_2$. Focusing on the propagating degree of freedom $\delp$, we find that its radial sound speed is 
\begin{equation}
  c_{\rho}^2 = -4 \frac{g_2(g_1^2 +\bar{G}_4g_2) }{g_1^2  \Lambda^8 \bar{P}_{,XX}}\frac{r_{\rm s}}{r^3} \,.
\label{csrho}
\end{equation}
As expected, this matches the result~\eqref{csho ell = 0} for~$\ell = 0$, since scalar modes with different~$\ell$ should have the same radial sound speed. The absence of ghosts and radial gradient stability~($c_{\rho}^2 > 0$) require
\be
 \bar{P}_{,XX} > 0\,; \qquad  g_2\left(g_1^2 +\bar{G}_4 g_2\right) < 0 \,.
\label{dipole stability}
\ee
Not surprisingly, these are identical respectively to the conditions~\eqref{ghost free} and~\eqref{radial stability} found in the monopole case. 

A key difference is the effective mass term, which can be read off from the last term in the first line of~\eqref{L dipole pi h2}:
\begin{equation}
 - \frac{1}{2} m^2_{\rm eff}(r) \delp^2=  -  2 \frac{g_2}{g_1^2} \left(g_1^2 +\bar{G}_4g_2\right) \left(\frac{r_{\rm s}}{r}\right)^{3/2} \frac{\delp^2}{r^2}  \,.
\label{m(r)}
\end{equation}
It follows from~\eqref{dipole stability} that $m^2_{\rm eff}(r) < 0$, and hence that dipole perturbations suffer from a tachyonic instability. 

However, one should keep in mind that the effective mass term originates from angular derivatives acting on~$\delp$, like the centrifugal term in the
radial wave equation. Indeed, we expect that the angular sound speed is related to the canonical mass term via 
\be
m_{\rm eff, \,canonical}^2(r) = \frac{\ell(\ell+1)}{r^2} c_{\theta,\varphi}^2\,.
\label{meff (r)}
\ee
With $\ell = 1$, we can read off from~\eqref{L dipole pi h2} after canonically normalizing~$\pi$ that 
\begin{equation}
c_{\theta,\varphi}^2 = 2\frac{g_2(g_1^2 +\bar{G}_4g_2) }{g_1^2  \Lambda^8 \bar{P}_{,XX}}\frac{r_{\rm s}}{r^3}  = - \frac{1}{2} c_{\rho}^2\,.
\label{cs angular}
\end{equation}
Since $c_{\rho}^2 =-2 c_{\theta,\varphi}^2 $, a gradient instability in either the radial or angular direction seems inevitable.

A word of caution is necessary, however, since to rigorously establish a gradient instability in the angular directions would require proving that~$c_{\theta,\varphi}$ is indeed
independent of multipole moments and given by~\eqref{cs angular} for all~$\ell$. This seems plausible because, on the one hand, we have shown that a gradient instability in either the radial or angular direction is indeed inevitable at least in the decoupling limit. On the other hand, as already mentioned at the end of Sec.~\ref{sec:bgdsclrpert}, although $c_{\rho}^2$ and $c_{\theta,\varphi}^2$ independently receive gravitational corrections, the relation $c_{\rho}^2 =-2 c_{\theta,\varphi}^2$ is maintained with the same proportionality constant. This gives credence to the expectation that this relation, and the gradient instability it entails, is maintained for higher~$\ell$ as well. A rigorous proof of this statement will require completing the general analysis of Sec.~\ref{general ell sec}, which we leave for future work. 

What the dipole analysis unambiguously shows is that the parity-even $\ell = 1$ sector suffers at the very least from a tachyonic instability. This indicates that the hairy black hole solution is not the correct background about which to perturb. If the instability is promoted to a gradient instability, as discussed above, this would have the more fatal implication that the hairy black hole solution lies outside the regime of validity of the effective theory.

\subsection{General multipoles: preliminary results}
\label{general ell sec}

For completeness, in this Section we present a partial treatment of perturbations with arbitrary multipoles. Because the analysis is considerably more complex in the general case, we can only provide an incomplete analysis. 

The most convenient gauge we have found in the general case is
\be
H_0=K=Q=0 \,,
\ee
leaving us with five component fields: $H_1$, $H_2$, $\alpha$, $\beta$ and $\pi$. Up to a total derivative, the quadratic Lagrangian is
\bea
\nonumber
 {\cal L}^{(\ell)}_{\rm even} & =&   \frac{\Lambda^8}{2} \bar{P}_{,XX}  \sqrt{r_{\rm s} r^3} \dot{\pi}^2 + 2\big(\bar{G}_4-2g_1-g_2\big) \left(\frac{r_{\rm s}}{r}\right)^{3/2} \left(\frac{r}{r_{\rm s}}\delp'^2  - \frac{\ell(\ell+1) }{2r^2} \delp^2\right)  + \frac{\bar{G}_4}{2}  \sqrt{\frac{r_{\rm s}}{r} } H_2^2 \\
     &+&  H_2 \Bigg[  2(g_1+g_2)r_{\rm s} \ddot{\delp} - 2(\bar{G}_4-g_1) \left( \sqrt{\frac{r_{\rm s}}{r}} \delp' -
     \ell(\ell+1) \frac{r_s}{r^2} \delp \right)  
    + g_1 \frac{\ell(\ell+1) }{r}\left( \sqrt{\frac{r_{\rm s}}{r}} \left( r\alpha \right)^{\cdot}  -     \beta\right) \Bigg] \nonumber \\ 
&+& \frac{1}{2}  \ell(\ell+1) g_1  \sqrt{\frac{r}{r_{\rm s}} } H_1 ^2 + H_1 \Bigg[2g_1 r \dot{H_2} - 4(g_1+g_2)r \dot{\pi}'  -g_1\ell(\ell+1) \sqrt{\frac{r}{r_{\rm s}} }\left(\alpha'  + \frac{\left(r \beta \right)^{\cdot}}{r}\right) \Bigg] \nonumber \\
  &+& \frac{1}{2}  \ell(\ell+1) g_1 \sqrt{\frac{r}{r_{\rm s}}}   \alpha'^2 + \frac{\ell(\ell + 1)}{r^2}g_1 \alpha \left[-\frac{g_1+g_2}{g_1} r_{\rm s} \dot{\pi} +  \frac{1}{\sqrt{r_{\rm s}}}\left( r^{5/2} \dot{\beta} \right)'   +  2 \left( r\beta \right)'   \right] \nonumber \\
  &+& \frac{1}{2}g_1 \sqrt{\frac{r}{r_{\rm s}} }  \dot{\beta}^2 +   2(\bar{G}_4-g_1) \ell(\ell+1) \beta\left( \frac{\delp}{r} \right)'\,,
\label{eqn:quadraticaction}
\eea
where, as before, implicit $(\ell,m)$ indices are summed over. Ignoring $\beta$, we see that this correctly matches the Lagrangians~\eqref{Levenl=0} and~\eqref{Levenl=1} with~$\ell = 0$ and~$\ell = 1$, respectively. Once again~$H_1$ is non-dynamical, and its equation of motion fixes its value to
\be
H_1 =  \frac{2}{\ell(\ell + 1)} \sqrt{\frac{r_{\rm s}}{r}} \left(-r \dot{H}_2 + 2 \frac{g_1 + g_2}{g_1} r\dot{\pi}' \right) + \alpha' +  \frac{\left(r \beta \right)^{\cdot}}{r}\,.
\ee
Substituting into~\eqref{Levenl=1} and integrating by parts,~$\alpha$ becomes a Lagrange multiplier, imposing the constraint
\bea
\nonumber
r \dot{H}_2' + \left(\frac{\ell(\ell + 1)}{2} + 1\right) \sqrt{\frac{r_{\rm s}}{r}} \dot{H}_2+ \frac{3}{4} \ell(\ell + 1) \frac{ r_{\rm s}}{r^2} H_2  &=&  \frac{\ell(\ell+1)}{2} \left( 2  \sqrt{\frac{r}{r_{\rm s}}} \dot{\beta}' + \frac{3}{r} \dot{\beta} + \frac{\beta'}{r}  + 3 \frac{\sqrt{r_{\rm s}}}{r^{5/2}} \beta \right) \\
&+& \frac{g_1 + g_2}{g_1}\left(2\left( r\dot{\pi}'   \right)' - \ell(\ell+1) \frac{r_{\rm s} }{r^2} \dot{\pi}\right) \,.
\label{alpha cons gen ell}
\eea
Ignoring the~$\beta$ terms, this matches~\eqref{alpha cons 1} with $\ell = 1$.

On the face of it,~\eqref{alpha cons gen ell} is quite a complicated constraint. However, there exists a convenient field definition,
\bea
\nonumber
\beta &\equiv & B + \frac{2 r}{\ell(\ell+1)}\left( \left( r H_2 \right)' - \frac{g_1+g_2}{g_1} \sqrt{r_{\rm s} r} \left( \frac{\chi}{r} \right)'\right) \,; \\
\pi  &=& \chi - \frac{g_1}{g_1 +g_2}  \sqrt{\frac{r^3}{r_{\rm s}}} H_2\,,
\label{B chi def}
\eea
which removes all terms containing derivatives of $H_2$. The constraint is then solved by
\be
H_2 = - \frac{1}{3} \frac{g_1+g_2}{g_1} \frac{\dot{\chi}}{r} + \frac{\ell(\ell + 1)}{\ell(\ell+1)-2} \frac{r}{r_{\rm s}} \left( \frac{2}{3}  \sqrt{\frac{r}{r_{\rm s}}} \dot{B}' + \frac{\dot{B} }{r} + \frac{B'}{3r} +  \frac{\sqrt{r_{\rm s}}}{r^{5/2}} B\right)\,.
\label{eqn:alphaconstraint}
\ee
Substitution into~\eqref{B chi def} allows one to express $\beta$ and $\pi$ as linear combinations of $\chi$, $\dot{\chi}$, $B$, $\dot{B}$ and their spatial derivatives, which therefore represents an invertible field redefinition. 

Inserting all these quantities back into~\eqref{eqn:quadraticaction}, one obtains a Lagrangian density in terms of only two variables, $\chi$ and $B$, albeit including higher time-derivative terms, such as $\ddot{\chi}^2$ and $\ddot{B}^2$. Nevertheless this should only describe two physically-propagating degrees of freedom. Hence, we expect there ought to exist a suitable field redefinition, for instance involving a linear combination of  $\dot{B}'$, $\dot{B}$, $B$, $\dot{\chi}$ and $\chi$, that would make this manifest. However this is technically challenging, and we have not been able to explicitly find the desired change of variables.

\section{Discussion}

General scalar tensor theories allow for a host of new approaches to the problems of modern cosmology. Perhaps the most fundamental theoretical constraint one can put on such theories is that they be ghost-free, and the clearest way to guarantee this is to restrict to second-order equations of motion. The resulting models --- Horndeski theories and their generalizations --- admit a rich phenomenology in general, and have been exploited for applications to both the early and the late universe. A particularly interesting subclass of these theories that has proven to have interesting cosmological implications consists of shift-symmetric theories. These models admit solutions that break time-translation invariance in a simple and interesting way, and as a byproduct, they evade a number of established black hole theorems in the literature. Their hairy black hole solutions provide both an interesting playground for constraining these theories through observational tests, and the possibility of new theoretical problems that allow us to further shrink the space of allowable models.

In this paper, we have studied the stability of a class of such hairy black hole solutions, and have identified a fatal instability. Taken together with previous results, our analysis allows a strong statement, that non-trivial black hole solutions with $\bar{X}={\rm constant} \neq 0$ in Horndeski theories are ruled out. These classical solutions are either unstable, or the effective field theory of perturbations around them is strongly coupled and cannot be trusted. 

This result complements earlier studies~\cite{Creminelli:2016zwa} of cosmological solutions and wormholes, especially of alternatives to inflation, in which other problems of Horndeski theories have been identified. 

We can think of at least two ways in which the result in this paper might be evaded by changing some of our key assumptions. One possibility is that the types of time-dependent hairy solutions that we consider might be stable in so-called beyond Horndeski or DHOST theories. Another possibility would be to consider the Horndeski terms as a subset of the operators allowed in a full effective field theory treatment (see, for example,~\cite{Solomon:2017nlh}), and to search for an effective operator that stabilizes the perturbations. The effective field theory (EFT) for quasinormal modes of a spherically-symmetric space-time with a scalar field inheriting the symmetry of the space-time is derived in~\cite{Franciolini:2018uyq}. The essential point is that the scalar field is only a function of the radial coordinate~$r$, such that one can choose a space-time slicing in which different constant values of $\phi$ define the slicing in $r$  (analogous to unitary gauge in the EFT of inflation). Thus, from the EFT it is not hard to identity the corresponding operators that stabilize the throat of a wormhole~\cite{Franciolini:2018aad}, for example. However in the case we are considering, the presence of a time-dependent scalar makes for a much more complicated system. 

In future work we will attempt to complete the analysis of Sec.~\ref{general ell sec} for even-sector perturbations with arbitrary~$\ell$. Furthermore, we will generalize our analysis to the case of asymptotically de Sitter and anti-de Sitter black hole solutions.

\bigskip
\goodbreak
\centerline{\bf Acknowledgements}

We thank Austin Joyce, Toshifumi Noumi, Enrico Trincherini, and especially Luca Santoni for useful discussions. This work is supported in part by US Department of Energy (HEP) Award DE-SC0013528. The work of J.K. and M.T. is also supported by NASA ATP grant 80NSSC18K0694, and by the Simons Foundation Origins of the Universe Initiative, grant number 658904.

\appendix 
\section{Lema\^{i}tre coordinates for static, spherically-symmetric space-times} \label{sec:AppdxA}

Lema\^{i}tre-type coordinate systems are examples of synchronous coordinate systems, in the sense that the global time of the metric matches the comoving time~$\tau$ of the observer:
\begin{equation}
   g_{\tau \tau} =-1\,.
\end{equation}
We start from the fact that, in general, a static, spherically-symmetric space-time can be written in Schwarzschild-type coordinates as
\begin{equation}
  \rd s^2  = - f(r) \rd t^2 + \frac{\rd r^2}{g(r)}  + r^2 \rd \Omega^2\,.
\end{equation}
Lema\^{i}tre-type coordinates for this metric are then given by
\begin{equation}
  \rd s^2  = -  \rd \tau ^2 + \big(1- f(r) \big)\rd \rho^2 + r^2 \rd \Omega^2\,,
\end{equation}
with
\begin{equation}
   \rd \tau = \rd t + \sqrt{\frac{1-f(r)}{f(r) g(r)}} \rd r\,;\qquad  \rd \rho = \rd t + \frac{1}{\sqrt{\big(1-f(r)\big) f(r) g(r) }} \rd r\,.
\end{equation}

As a special case, the Schwarzschild metric, with $f(r) = g(r) = 1 - \frac{r_{\rm s}}{r}$, becomes in Lema\^{i}tre coordinates
\be
\rd s^2 = -\rd\tau^2 + \frac{r_{\rm s}}{r} \rd \rho^2 + r^2 \rd\Omega_2^2\,,
\ee
with
\be
 \rd \tau = \rd t + \frac{\sqrt{r_{\rm s} r}}{r-r_{\rm s}}  \rd r\,; \qquad  \rd \rho = \rd t + \sqrt{\frac{r^3}{r_{\rm s}}} \frac{1}{r-r_{\rm s}}\rd r\,.
\ee
These integrate to 
\be
\tau = t + 2\sqrt{r r_{\rm s}} + r_{\rm s} \ln\left| \frac{\sqrt{r} - \sqrt{r_{\rm s}} }{\sqrt{r} + \sqrt{r_{\rm s}}}\right| \,;\qquad \rho = \tau + \frac{2}{3} r_{\rm s} \left(\frac{r}{r_{\rm s}}\right)^{3/2}\,,
\label{taurho}
\ee
which imply
\begin{equation}  
  \qquad  r= \left[ \frac{3}{2}(\rho-\tau)\right]^{2/3} r_{\rm s}^{1/3}\,.
\end{equation}
Trajectories of constant~$\rho$ are time-like geodesics, freely-falling into the black hole and ultimately hitting the singularity at~$\rho-\tau=0$. Note also that the Lema\^{i}tre coordinates cover half of the maximally-extended Schwarzschild black hole region (the ``in-going" regions I and II)~\cite{Mukohyama:2005rw}.

\section{Scalar-vector-tensor spherical harmonics} 
\label{sec:appdxB}

The building blocks of the scalar-vector-tensor (SVT) decomposition on the two-sphere are the familiar scalar spherical harmonics, $Y_\ell^m(\theta,\varphi)$, assumed to be real-valued. 
They are defined as usual as eigenfunctions of the angular Laplacian,
\be
\label{eqn:hamonics}
 g^{AB}\nabla_A \nabla_B Y_\ell^m(\theta,\varphi) = - \ell (\ell +1)Y_\ell^m(\theta,\varphi)\,,
\ee
where $g_{AB}$ and $\nabla_A$ are respectively the metric and covariant derivative on the two-sphere, with $A,B$ indices denoting angular coordinates $\theta$ and $\varphi$.
The spherical harmonic $Y_\ell^m(\theta,\varphi)$ has parity eigenvalue $(-1)^\ell$, and the spherical harmonics satisfy the orthonormality relation:
\be
\int \rd  \Omega   \,Y_\ell^m(\theta,\varphi) Y_{\ell'}^{m'}(\theta,\varphi)=  \delta_{m m'} \delta_{\ell \ell'} \,.
\label{Ylm ortho}
\ee 

Vector and tensor spherical harmonics can be expressed as derivatives of the scalar harmonics $Y_\ell^m$. Our conventions are those of~\cite{Martel:2005ir}. 
Vector spherical harmonics can be decomposed into components of opposite parity, 
\be
Y_{A\,\ell}^{~\;\;m} (\theta,\varphi)= \nabla_A Y_\ell^m(\theta,\varphi) \,;\qquad X_{A\,\ell}^{~\;\;m}(\theta,\varphi) = \epsilon_A^{\;\;B} \nabla_B Y_\ell^m(\theta,\varphi)\,,
\ee
with parity $(-1)^\ell$ and $(-1)^{\ell + 1}$, respectively. It is customary in the literature to refer to $Y_{A\,\ell}^{~\;\;m}$'s as even (or electric-type) vector harmonics, and $X_{A\,\ell}^{~\;\;m}$ as odd (or magnetic-type) vector harmonics. The orthonormality condition~\eqref{Ylm ortho} implies
\bea
 \int \rd  \Omega   \, g^{AB} Y_{A\,\ell}^{~\;\;m}  Y_{B\,\ell'}^{~\;\;m'} & = & \ell(\ell+1) \delta_{m m'} \delta_{\ell \ell'}\,; \nonumber \\
  \int \rd  \Omega   \, g^{AB} X_{A\,\ell}^{~\;\;m}  X_{B\,\ell'}^{~\;\;m'}  &= &\ell(\ell+1) \delta_{m m'} \delta_{\ell \ell'}\,; \nonumber \\
   \int \rd  \Omega   \, g^{AB} X_{A\,\ell}^{~\;\;m}  Y_{B\,\ell'}^{~\;\;m'}  &=&  0\,.
\eea

Similarly, tensor spherical harmonics decompose into components of opposite parity,
\bea
\nonumber
Y_{AB\,\ell}^{~\;\;\;\;\;m} (\theta,\varphi) &=& \left(\nabla_A\nabla_B + \frac{1}{2}\ell(\ell+1) g_{AB} \right) Y_\ell^m(\theta,\varphi)\,; \\
X_{AB\,\ell}^{~\;\;\;\;\;\,m}(\theta,\varphi) &=&  \epsilon_{(A}^{\;\;\;C} \nabla_{B)} \nabla_C Y_\ell^m(\theta,\varphi)\,, 
\eea
with respective parity of $(-1)^\ell$ and $(-1)^{\ell + 1}$. As with vector harmonics, we refer to $Y_{AB\,\ell}^{~\;\;\;\;\;m}$ as even (or electric-type) tensor harmonics, and $X_{AB\,\ell}^{~\;\;\;\;\;\,m}$ as odd (or magnetic-type) tensor harmonics. The orthonormality relation~\eqref{Ylm ortho} in this case implies
\bea
 \int \rd  \Omega   \, g^{AC}g^{BD} Y_{AB\,\ell}^{~\;\;\;\;\;m}Y_{CD\,\ell'}^{~\;\;\;\;\;m'}  &=& \frac{1}{2}(\ell-1)\ell (\ell+1)(\ell+2) \delta_{m m'} \delta_{\ell \ell'} \,; \nonumber \\
  \int \rd  \Omega   \, g^{AC}g^{BD} X_{AB\,\ell}^{~\;\;\;\;\;\,m}X_{CD\,\ell'}^{~\;\;\;\;\;\,m'}   &=& \frac{1}{2}(\ell-1)\ell (\ell+1)(\ell+2) \delta_{m m'} \delta_{\ell \ell'} \,; \nonumber \\
    \int \rd  \Omega   \, g^{AC}g^{BD} X_{AB\,\ell}^{~\;\;\;\;\;\,m}Y_{CD\,\ell'}^{~\;\;\;\;\;\,m'} &=& 0\,.
\eea

\section{Contributions to the equations of motion from~${\cal L}_3$ and~${\cal L}_5$} \label{sec:L3L5eom}

In this Appendix we collect the remaining terms in the scalar equation of motion and Einstein's field equations for the shift-symmetric
Horndeski theory~\eqref{Horndeski} discussed in Sec.~\ref{BH soln arbitrary G4}. 

The scalar field equation of motion is given by
\be
{\cal E}^{\phi}_2+{\cal E}^{\phi}_3+{\cal E}^{\phi}_4+{\cal E}^{\phi}_5 =0\,.
\ee
with~${\cal E}^{\phi}_i$ and~$T_{i \mu\nu}$ derived from the corresponding~${\cal L}_i$ in~\eqref{L_i}. The contributions~${\cal E}^{\phi}_2$ and~${\cal E}^{\phi}_4$
are given in~\eqref{E2E4}. The explicit expressions for~${\cal E}^{\phi}_3$ and~${\cal E}^{\phi}_5$ are 
\bea
{\cal E}^{\phi}_3&=& \nabla_{\mu}\Big(G_{3,X} \big(\square \phi \nabla^{\mu}\phi + \nabla^{\mu} X\big) \Big) \,; \nonumber \\
{\cal E}^{\phi}_5&=& \nabla_{\mu} \Bigg( -6G_{5,X}\big( G^{\mu\nu} \nabla_{\nu} X + G^{\rho\sigma} \phi_{\rho\sigma} \nabla^{\mu} \phi \big) + G_{5,XX}\Big((\square \phi)^3 - 3 \square \phi \phi_{\rho\sigma} \phi^{\rho\sigma}+ 2 \phi^{\;\;\sigma}_{ \alpha}\phi^{\;\;\rho}_{ \sigma}\phi^{\;\;\alpha}_{ \rho} \Big) \nabla^{\mu}\phi  \nonumber \\
 && \qquad +~\nabla_{\nu}\Big( G_{5,X}\left[3 (\square \phi)^2 g^{\mu\nu} - 3\phi_{\rho\sigma} \phi^{\rho\sigma} g^{\mu\nu} - 6 \square \phi  \phi^{\mu\nu} + 6 \phi^{\mu}_{\;\; \rho} \phi^{\rho\nu}  \right] \Big)  \Bigg)\,.
\eea

For our black hole background, the Einstein field equations for the metric tensor require the vanishing of the total stress-energy tensor: 
\be
T_{2 \mu\nu}+T_{3 \mu\nu}+T_{4 \mu\nu}+T_{5 \mu\nu} =0\,.
\ee
The contributions~$T_{2 \mu\nu}$ and~$T_{4 \mu\nu}$ are given in~\eqref{T2T4}. The explicit expressions for~$T_{3 \mu\nu}$ and~$T_{5 \mu\nu}$ are
\bea
T_{3 \mu\nu} &=& G_{3,X}\Big(\square \phi \nabla_{\mu}\phi \nabla_{\nu} \phi -g_{\mu\nu} \nabla_{\rho} \phi \nabla^{\rho} X + 2 \nabla_{(\mu} \phi \nabla_{\nu)} X\Big)\,; \\
T_{5,\mu\nu} &=& \Bigg\{  G_{5,X} \Big(  R_{\rho\sigma} \big[ 6\nabla^{\rho}\phi \nabla^{\sigma} \phi \square \phi + 12 \nabla^{\rho}\phi \nabla^{\sigma} X \big]- 3R \nabla_{\rho}\phi \nabla^{\rho}X - 6 R_{\alpha \beta\rho\sigma} \nabla^{\alpha}\phi \nabla^{\rho} \phi^{\beta \sigma}     \nonumber \\
&& \qquad \qquad\;\; -2 \big[ (\square \phi)^3 - 3 \square \phi \phi_{\rho\sigma} \phi^{\rho\sigma}+ 2 \phi^{\;\;\sigma}_{ \alpha}\phi^{\;\;\rho}_{ \sigma}\phi^{\;\;\alpha}_{ \rho} \big]  \Big) \nonumber \\
&& \qquad +~G_{5,XX} \Big(6 \phi_{\rho\sigma} \nabla^{\rho}X \nabla^{\sigma}X - 6 \square \phi \nabla{\rho} X\nabla_{\rho}X -3 \left[ (\square \phi)^2 - \phi_{\rho\sigma} \phi^{\rho\sigma}\right]\nabla_{\alpha}\phi \nabla^{\alpha}X  \Big) \Bigg\} g_{\mu\nu}  \nonumber \\
&+&  \Bigg( G_{5,XX}\big[  (\square \phi)^3 - 3 \square \phi \phi_{\rho\sigma} \phi^{\rho\sigma}+ 2 \phi^{\;\;\sigma}_{ \alpha}\phi^{\;\;\rho}_{ \sigma}\phi^{\;\;\alpha}_{ \rho}\big] - 6G_{5,X} G_{\rho\sigma} \phi^{\rho\sigma} \Bigg) \nabla_{\mu}\phi \nabla_{\nu} \phi  \nonumber \\
&+& 6 G_{5,XX} \square \phi \nabla_{\mu}X\nabla_{\nu}X + \Bigg(6 G_{5,X} R + 6G_{5,XX} \big[ (\square \phi)^2 - \phi_{\rho\sigma} \phi^{\rho\sigma} \big] \Bigg) \nabla_{(\mu} \phi\nabla_{\nu)} X  \nonumber\\
&+& 6\Bigg(  G_{5,X}\big[(\square \phi)^2 -\phi_{\rho\sigma} \phi^{\rho\sigma} - R_{\rho\sigma} \nabla^{\rho}\phi \nabla^{\sigma}\phi \big]  + G_{5,XX}\big(\nabla_{\rho}X\nabla^{\rho}X + \square \phi \nabla_{\rho}\phi \nabla^{\rho}X \big) \Bigg) \phi_{\mu\nu}  
\nonumber \\
&+&12G_{5,XX} \bigg(\nabla_{\alpha} X \phi^{\alpha \beta} \phi_{\beta (\mu} \nabla_{\nu)}\phi  + \nabla^{\rho}X \phi_{\rho (\mu} \nabla_{\nu)}X - \square \phi \nabla^{\rho} X \phi_{\rho ( \mu} \nabla_{\nu)} \phi    \bigg) \nonumber \\
&-& \bigg( 12 G_{5,X}\square \phi + 6 G_{5,XX} \nabla_{\rho}\phi \nabla^{\rho}X \bigg) \phi_{\mu}^{\;\;\rho}\phi_{\rho \nu} +12 G_{5,X} \phi_{\alpha \beta} \phi^{\beta}_{\;\; \mu} \phi^{\alpha}_{\;\;\nu} \nonumber \\
& -& 6 G_{5,X} \Bigg( 2 (\square \phi \nabla^{\rho}\phi + \nabla^{\rho} X) R_{\rho(\mu} \nabla_{\nu)}\phi  - 2R_{\rho\sigma} \nabla^{\rho} \phi^{\sigma}_{\;\; (\mu} \nabla_{\nu)}\phi+ 2 \nabla^{\rho} R_{\rho (\mu} \nabla_{\nu)}X - \nabla_{\rho}\phi\nabla^{\rho}X R_{\mu\nu} \nonumber \\
& &\qquad \;\;\; +~2 \nabla_{\alpha}\phi \phi^{\rho\sigma} R^{\alpha}_{\;\; \rho\sigma (\mu} \nabla_{\nu)}\phi - R_{\rho(\mu\nu)\sigma}\nabla^{\rho}\phi(\square\phi \nabla^{\sigma} \phi + 2 \nabla^{\sigma}X)  -2 \nabla^{\rho} \phi \nabla^{\alpha} \phi R_{\alpha \sigma \rho (\mu} \phi^{\sigma}_{\;\; \nu)} \Bigg)\,. \nonumber\\
\eea

\renewcommand{\em}{}
\bibliographystyle{utphys}
\bibliography{Horndeski_BH_hair_v7.bib}

\providecommand{\href}[2]{#2}\begingroup\raggedright\begin{thebibliography}{10}

\bibitem{TheLIGOScientific:2016pea}
{\bf LIGO Scientific, Virgo} Collaboration, B.~Abbott {\em et al.}, ``{Binary
  Black Hole Mergers in the first Advanced LIGO Observing Run},''
  \href{http://dx.doi.org/10.1103/PhysRevX.6.041015}{{\em Phys.\ Rev.\ X} {\bf
  6} (2016) no.~4, 041015}, \href{http://arxiv.org/abs/1606.04856}{{\tt
  arXiv:1606.04856 [gr-qc]}}. [Erratum: Phys.Rev.X 8, 039903 (2018)].

\bibitem{Akiyama:2019cqa}
{\bf Event Horizon Telescope} Collaboration, K.~Akiyama {\em et al.}, ``{First
  M87 Event Horizon Telescope Results. I. The Shadow of the Supermassive Black
  Hole},'' \href{http://dx.doi.org/10.3847/2041-8213/ab0ec7}{{\em Astrophys.\
  J.} {\bf 875} (2019) no.~1, L1}, \href{http://arxiv.org/abs/1906.11238}{{\tt
  arXiv:1906.11238 [astro-ph.GA]}}.

\bibitem{Berti:2009kk}
E.~Berti, V.~Cardoso, and A.~O. Starinets, ``{Quasinormal modes of black holes
  and black branes},''
  \href{http://dx.doi.org/10.1088/0264-9381/26/16/163001}{{\em Class.\ Quant.\
  Grav.} {\bf 26} (2009)  163001}, \href{http://arxiv.org/abs/0905.2975}{{\tt
  arXiv:0905.2975 [gr-qc]}}.

\bibitem{Woodard:2015zca}
R.~P. Woodard, ``{Ostrogradsky's theorem on Hamiltonian instability},''
  \href{http://dx.doi.org/10.4249/scholarpedia.32243}{{\em Scholarpedia} {\bf
  10} (2015) no.~8, 32243}, \href{http://arxiv.org/abs/1506.02210}{{\tt
  arXiv:1506.02210 [hep-th]}}.

\bibitem{Horndeski:1974wa}
G.~W. Horndeski, ``{Second-order scalar-tensor field equations in a
  four-dimensional space},'' \href{http://dx.doi.org/10.1007/BF01807638}{{\em
  Int.\ J.\ Theor.\ Phys.} {\bf 10} (1974)  363--384}.

\bibitem{Deffayet:2009mn}
C.~Deffayet, S.~Deser, and G.~Esposito-Farese, ``{Generalized Galileons: All
  scalar models whose curved background extensions maintain second-order field
  equations and stress-tensors},''
  \href{http://dx.doi.org/10.1103/PhysRevD.80.064015}{{\em Phys. Rev. D} {\bf
  80} (2009)  064015}, \href{http://arxiv.org/abs/0906.1967}{{\tt
  arXiv:0906.1967 [gr-qc]}}.

\bibitem{Deffayet:2011gz}
C.~Deffayet, X.~Gao, D.~Steer, and G.~Zahariade, ``{From k-essence to
  generalised Galileons},''
  \href{http://dx.doi.org/10.1103/PhysRevD.84.064039}{{\em Phys. Rev. D} {\bf
  84} (2011)  064039}, \href{http://arxiv.org/abs/1103.3260}{{\tt
  arXiv:1103.3260 [hep-th]}}.

\bibitem{Kobayashi:2011nu}
T.~Kobayashi, M.~Yamaguchi, and J.~Yokoyama, ``{Generalized G-inflation:
  Inflation with the most general second-order field equations},''
  \href{http://dx.doi.org/10.1143/PTP.126.511}{{\em Prog. Theor. Phys.} {\bf
  126} (2011)  511--529}, \href{http://arxiv.org/abs/1105.5723}{{\tt
  arXiv:1105.5723 [hep-th]}}.

\bibitem{Gleyzes:2014dya}
J.~Gleyzes, D.~Langlois, F.~Piazza, and F.~Vernizzi, ``{Healthy theories beyond
  Horndeski},'' \href{http://dx.doi.org/10.1103/PhysRevLett.114.211101}{{\em
  Phys. Rev. Lett.} {\bf 114} (2015) no.~21, 211101},
  \href{http://arxiv.org/abs/1404.6495}{{\tt arXiv:1404.6495 [hep-th]}}.

\bibitem{Langlois:2015cwa}
D.~Langlois and K.~Noui, ``{Degenerate higher derivative theories beyond
  Horndeski: evading the Ostrogradski instability},''
  \href{http://dx.doi.org/10.1088/1475-7516/2016/02/034}{{\em JCAP} {\bf 02}
  (2016)  034}, \href{http://arxiv.org/abs/1510.06930}{{\tt arXiv:1510.06930
  [gr-qc]}}.

\bibitem{Crisostomi:2016czh}
M.~Crisostomi, K.~Koyama, and G.~Tasinato, ``{Extended Scalar-Tensor Theories
  of Gravity},'' \href{http://dx.doi.org/10.1088/1475-7516/2016/04/044}{{\em
  JCAP} {\bf 04} (2016)  044}, \href{http://arxiv.org/abs/1602.03119}{{\tt
  arXiv:1602.03119 [hep-th]}}.

\bibitem{BenAchour:2016fzp}
J.~Ben~Achour, M.~Crisostomi, K.~Koyama, D.~Langlois, K.~Noui, and G.~Tasinato,
  ``{Degenerate higher order scalar-tensor theories beyond Horndeski up to
  cubic order},'' \href{http://dx.doi.org/10.1007/JHEP12(2016)100}{{\em JHEP}
  {\bf 12} (2016)  100}, \href{http://arxiv.org/abs/1608.08135}{{\tt
  arXiv:1608.08135 [hep-th]}}.

\bibitem{Takahashi:2017pje}
K.~Takahashi and T.~Kobayashi, ``{Extended mimetic gravity: Hamiltonian
  analysis and gradient instabilities},''
  \href{http://dx.doi.org/10.1088/1475-7516/2017/11/038}{{\em JCAP} {\bf 11}
  (2017)  038}, \href{http://arxiv.org/abs/1708.02951}{{\tt arXiv:1708.02951
  [gr-qc]}}.

\bibitem{Langlois:2018jdg}
D.~Langlois, M.~Mancarella, K.~Noui, and F.~Vernizzi, ``{Mimetic gravity as
  DHOST theories},''
  \href{http://dx.doi.org/10.1088/1475-7516/2019/02/036}{{\em JCAP} {\bf 02}
  (2019)  036}, \href{http://arxiv.org/abs/1802.03394}{{\tt arXiv:1802.03394
  [gr-qc]}}.

\bibitem{Sotiriou:2013qea}
T.~P. Sotiriou and S.-Y. Zhou, ``{Black hole hair in generalized scalar-tensor
  gravity},'' \href{http://dx.doi.org/10.1103/PhysRevLett.112.251102}{{\em
  Phys. Rev. Lett.} {\bf 112} (2014)  251102},
  \href{http://arxiv.org/abs/1312.3622}{{\tt arXiv:1312.3622 [gr-qc]}}.

\bibitem{Sotiriou:2014pfa}
T.~P. Sotiriou and S.-Y. Zhou, ``{Black hole hair in generalized scalar-tensor
  gravity: An explicit example},''
  \href{http://dx.doi.org/10.1103/PhysRevD.90.124063}{{\em Phys. Rev. D} {\bf
  90} (2014)  124063}, \href{http://arxiv.org/abs/1408.1698}{{\tt
  arXiv:1408.1698 [gr-qc]}}.

\bibitem{Babichev:2016rlq}
E.~Babichev, C.~Charmousis, and A.~Lehébel, ``{Black holes and stars in
  Horndeski theory},''
  \href{http://dx.doi.org/10.1088/0264-9381/33/15/154002}{{\em Class. Quant.
  Grav.} {\bf 33} (2016) no.~15, 154002},
  \href{http://arxiv.org/abs/1604.06402}{{\tt arXiv:1604.06402 [gr-qc]}}.

\bibitem{Benkel:2016rlz}
R.~Benkel, T.~P. Sotiriou, and H.~Witek, ``{Black hole hair formation in
  shift-symmetric generalised scalar-tensor gravity},''
  \href{http://dx.doi.org/10.1088/1361-6382/aa5ce7}{{\em Class. Quant. Grav.}
  {\bf 34} (2017) no.~6, 064001}, \href{http://arxiv.org/abs/1610.09168}{{\tt
  arXiv:1610.09168 [gr-qc]}}.

\bibitem{Babichev:2017guv}
E.~Babichev, C.~Charmousis, and A.~Lehébel, ``{Asymptotically flat black holes
  in Horndeski theory and beyond},''
  \href{http://dx.doi.org/10.1088/1475-7516/2017/04/027}{{\em JCAP} {\bf 04}
  (2017)  027}, \href{http://arxiv.org/abs/1702.01938}{{\tt arXiv:1702.01938
  [gr-qc]}}.

\bibitem{Lehebel:2017fag}
A.~Lehébel, E.~Babichev, and C.~Charmousis, ``{A no-hair theorem for stars in
  Horndeski theories},''
  \href{http://dx.doi.org/10.1088/1475-7516/2017/07/037}{{\em JCAP} {\bf 07}
  (2017)  037}, \href{http://arxiv.org/abs/1706.04989}{{\tt arXiv:1706.04989
  [gr-qc]}}.

\bibitem{Minamitsuji:2018vuw}
M.~Minamitsuji and H.~Motohashi, ``{Stealth Schwarzschild solution in shift
  symmetry breaking theories},''
  \href{http://dx.doi.org/10.1103/PhysRevD.98.084027}{{\em Phys. Rev. D} {\bf
  98} (2018) no.~8, 084027}, \href{http://arxiv.org/abs/1809.06611}{{\tt
  arXiv:1809.06611 [gr-qc]}}.

\bibitem{BenAchour:2019fdf}
J.~Ben~Achour, H.~Liu, and S.~Mukohyama, ``{Hairy black holes in DHOST
  theories: Exploring disformal transformation as a solution-generating
  method},'' \href{http://dx.doi.org/10.1088/1475-7516/2020/02/023}{{\em JCAP}
  {\bf 02} (2020)  023}, \href{http://arxiv.org/abs/1910.11017}{{\tt
  arXiv:1910.11017 [gr-qc]}}.

\bibitem{Minamitsuji:2019tet}
M.~Minamitsuji and J.~Edholm, ``{Black holes with a nonconstant kinetic term in
  degenerate higher-order scalar tensor theories},''
  \href{http://dx.doi.org/10.1103/PhysRevD.101.044034}{{\em Phys. Rev. D} {\bf
  101} (2020) no.~4, 044034}, \href{http://arxiv.org/abs/1912.01744}{{\tt
  arXiv:1912.01744 [gr-qc]}}.

\bibitem{Kobayashi:2012kh}
T.~Kobayashi, H.~Motohashi, and T.~Suyama, ``{Black hole perturbation in the
  most general scalar-tensor theory with second-order field equations I: the
  odd-parity sector},''
  \href{http://dx.doi.org/10.1103/PhysRevD.85.084025}{{\em Phys.\ Rev.\ D} {\bf
  85} (2012)  084025}, \href{http://arxiv.org/abs/1202.4893}{{\tt
  arXiv:1202.4893 [gr-qc]}}. [Erratum: Phys.Rev.D 96, 109903 (2017)].

\bibitem{Kobayashi:2014wsa}
T.~Kobayashi, H.~Motohashi, and T.~Suyama, ``{Black hole perturbation in the
  most general scalar-tensor theory with second-order field equations II: the
  even-parity sector},''
  \href{http://dx.doi.org/10.1103/PhysRevD.89.084042}{{\em Phys.\ Rev.\ D} {\bf
  89} (2014) no.~8, 084042}, \href{http://arxiv.org/abs/1402.6740}{{\tt
  arXiv:1402.6740 [gr-qc]}}.

\bibitem{Franciolini:2018uyq}
G.~Franciolini, L.~Hui, R.~Penco, L.~Santoni, and E.~Trincherini, ``{Effective
  Field Theory of Black Hole Quasinormal Modes in Scalar-Tensor Theories},''
  \href{http://dx.doi.org/10.1007/JHEP02(2019)127}{{\em JHEP} {\bf 02} (2019)
  127}, \href{http://arxiv.org/abs/1810.07706}{{\tt arXiv:1810.07706
  [hep-th]}}.

\bibitem{Babichev:2013cya}
E.~Babichev and C.~Charmousis, ``{Dressing a black hole with a time-dependent
  Galileon},'' \href{http://dx.doi.org/10.1007/JHEP08(2014)106}{{\em JHEP} {\bf
  08} (2014)  106}, \href{http://arxiv.org/abs/1312.3204}{{\tt arXiv:1312.3204
  [gr-qc]}}.

\bibitem{Kobayashi:2014eva}
T.~Kobayashi and N.~Tanahashi, ``{Exact black hole solutions in shift symmetric
  scalar--tensor theories},'' \href{http://dx.doi.org/10.1093/ptep/ptu096}{{\em
  PTEP} {\bf 2014} (2014)  073E02}, \href{http://arxiv.org/abs/1403.4364}{{\tt
  arXiv:1403.4364 [gr-qc]}}.

\bibitem{Babichev:2016kdt}
E.~Babichev and G.~Esposito-Farese, ``{Cosmological self-tuning and local
  solutions in generalized Horndeski theories},''
  \href{http://dx.doi.org/10.1103/PhysRevD.95.024020}{{\em Phys. Rev. D} {\bf
  95} (2017) no.~2, 024020}, \href{http://arxiv.org/abs/1609.09798}{{\tt
  arXiv:1609.09798 [gr-qc]}}.

\bibitem{Babichev:2017lmw}
E.~Babichev, C.~Charmousis, G.~Esposito-Farèse, and A.~Lehébel, ``{Stability
  of Black Holes and the Speed of Gravitational Waves within Self-Tuning
  Cosmological Models},''
  \href{http://dx.doi.org/10.1103/PhysRevLett.120.241101}{{\em Phys. Rev.
  Lett.} {\bf 120} (2018) no.~24, 241101},
  \href{http://arxiv.org/abs/1712.04398}{{\tt arXiv:1712.04398 [gr-qc]}}.

\bibitem{BenAchour:2018dap}
J.~Ben~Achour and H.~Liu, ``{Hairy Schwarzschild-(A)dS black hole solutions in
  degenerate higher order scalar-tensor theories beyond shift symmetry},''
  \href{http://dx.doi.org/10.1103/PhysRevD.99.064042}{{\em Phys. Rev. D} {\bf
  99} (2019) no.~6, 064042}, \href{http://arxiv.org/abs/1811.05369}{{\tt
  arXiv:1811.05369 [gr-qc]}}.

\bibitem{Motohashi:2019sen}
H.~Motohashi and M.~Minamitsuji, ``{Exact black hole solutions in
  shift-symmetric quadratic degenerate higher-order scalar-tensor theories},''
  \href{http://dx.doi.org/10.1103/PhysRevD.99.064040}{{\em Phys.\ Rev.\ D} {\bf
  99} (2019) no.~6, 064040}, \href{http://arxiv.org/abs/1901.04658}{{\tt
  arXiv:1901.04658 [gr-qc]}}.

\bibitem{Takahashi:2019oxz}
K.~Takahashi, H.~Motohashi, and M.~Minamitsuji, ``{Linear stability analysis of
  hairy black holes in quadratic degenerate higher-order scalar-tensor
  theories: Odd-parity perturbations},''
  \href{http://dx.doi.org/10.1103/PhysRevD.100.024041}{{\em Phys.\ Rev.\ D}
  {\bf 100} (2019) no.~2, 024041}, \href{http://arxiv.org/abs/1904.03554}{{\tt
  arXiv:1904.03554 [gr-qc]}}.

\bibitem{Minamitsuji:2019shy}
M.~Minamitsuji and J.~Edholm, ``{Black hole solutions in shift-symmetric
  degenerate higher-order scalar-tensor theories},''
  \href{http://dx.doi.org/10.1103/PhysRevD.100.044053}{{\em Phys. Rev. D} {\bf
  100} (2019) no.~4, 044053}, \href{http://arxiv.org/abs/1907.02072}{{\tt
  arXiv:1907.02072 [gr-qc]}}.

\bibitem{Mukohyama:2005rw}
S.~Mukohyama, ``{Black holes in the ghost condensate},''
  \href{http://dx.doi.org/10.1103/PhysRevD.71.104019}{{\em Phys.\ Rev.\ D} {\bf
  71} (2005)  104019}, \href{http://arxiv.org/abs/hep-th/0502189}{{\tt
  arXiv:hep-th/0502189}}.

\bibitem{ArkaniHamed:2003uz}
N.~Arkani-Hamed, P.~Creminelli, S.~Mukohyama, and M.~Zaldarriaga, ``{Ghost
  inflation},'' \href{http://dx.doi.org/10.1088/1475-7516/2004/04/001}{{\em
  JCAP} {\bf 04} (2004)  001}, \href{http://arxiv.org/abs/hep-th/0312100}{{\tt
  arXiv:hep-th/0312100}}.

\bibitem{Hui:2012qt}
L.~Hui and A.~Nicolis, ``{No-Hair Theorem for the Galileon},''
  \href{http://dx.doi.org/10.1103/PhysRevLett.110.241104}{{\em Phys.\ Rev.\
  Lett.} {\bf 110} (2013)  241104}, \href{http://arxiv.org/abs/1202.1296}{{\tt
  arXiv:1202.1296 [hep-th]}}.

\bibitem{Ogawa:2015pea}
H.~Ogawa, T.~Kobayashi, and T.~Suyama, ``{Instability of hairy black holes in
  shift-symmetric Horndeski theories},''
  \href{http://dx.doi.org/10.1103/PhysRevD.93.064078}{{\em Phys. Rev. D} {\bf
  93} (2016) no.~6, 064078}, \href{http://arxiv.org/abs/1510.07400}{{\tt
  arXiv:1510.07400 [gr-qc]}}.

\bibitem{Babichev:2018uiw}
E.~Babichev, C.~Charmousis, G.~Esposito-Farèse, and A.~Lehébel,
  ``{Hamiltonian unboundedness vs stability with an application to Horndeski
  theory},'' \href{http://dx.doi.org/10.1103/PhysRevD.98.104050}{{\em Phys.
  Rev. D} {\bf 98} (2018) no.~10, 104050},
  \href{http://arxiv.org/abs/1803.11444}{{\tt arXiv:1803.11444 [gr-qc]}}.

\bibitem{deRham:2019gha}
C.~de~Rham and J.~Zhang, ``{Perturbations of stealth black holes in degenerate
  higher-order scalar-tensor theories},''
  \href{http://dx.doi.org/10.1103/PhysRevD.100.124023}{{\em Phys.\ Rev.\ D}
  {\bf 100} (2019) no.~12, 124023}, \href{http://arxiv.org/abs/1907.00699}{{\tt
  arXiv:1907.00699 [hep-th]}}.

\bibitem{ArkaniHamed:2003uy}
N.~Arkani-Hamed, H.-C. Cheng, M.~A. Luty, and S.~Mukohyama, ``{Ghost
  condensation and a consistent infrared modification of gravity},''
  \href{http://dx.doi.org/10.1088/1126-6708/2004/05/074}{{\em JHEP} {\bf 05}
  (2004)  074}, \href{http://arxiv.org/abs/hep-th/0312099}{{\tt
  arXiv:hep-th/0312099}}.

\bibitem{Chen:2006nt}
X.~Chen, M.-x. Huang, S.~Kachru, and G.~Shiu, ``{Observational signatures and
  non-Gaussianities of general single field inflation},''
  \href{http://dx.doi.org/10.1088/1475-7516/2007/01/002}{{\em JCAP} {\bf 01}
  (2007)  002}, \href{http://arxiv.org/abs/hep-th/0605045}{{\tt
  arXiv:hep-th/0605045}}.

\bibitem{ArkaniHamed:2005gu}
N.~Arkani-Hamed, H.-C. Cheng, M.~A. Luty, S.~Mukohyama, and T.~Wiseman,
  ``{Dynamics of gravity in a Higgs phase},''
  \href{http://dx.doi.org/10.1088/1126-6708/2007/01/036}{{\em JHEP} {\bf 01}
  (2007)  036}, \href{http://arxiv.org/abs/hep-ph/0507120}{{\tt
  arXiv:hep-ph/0507120}}.

\bibitem{Wald:1993nt}
R.~M. Wald, ``{Black hole entropy is the Noether charge},''
  \href{http://dx.doi.org/10.1103/PhysRevD.48.R3427}{{\em Phys.\ Rev.\ D} {\bf
  48} (1993) no.~8, 3427--3431}, \href{http://arxiv.org/abs/gr-qc/9307038}{{\tt
  arXiv:gr-qc/9307038}}.

\bibitem{Cheung:2018cwt}
C.~Cheung, J.~Liu, and G.~N. Remmen, ``{Proof of the Weak Gravity Conjecture
  from Black Hole Entropy},''
  \href{http://dx.doi.org/10.1007/JHEP10(2018)004}{{\em JHEP} {\bf 10} (2018)
  004}, \href{http://arxiv.org/abs/1801.08546}{{\tt arXiv:1801.08546
  [hep-th]}}.

\bibitem{Hamada:2018dde}
Y.~Hamada, T.~Noumi, and G.~Shiu, ``{Weak Gravity Conjecture from Unitarity and
  Causality},'' \href{http://dx.doi.org/10.1103/PhysRevLett.123.051601}{{\em
  Phys.\ Rev.\ Lett.} {\bf 123} (2019) no.~5, 051601},
  \href{http://arxiv.org/abs/1810.03637}{{\tt arXiv:1810.03637 [hep-th]}}.

\bibitem{Regge:1957td}
T.~Regge and J.~A. Wheeler, ``{Stability of a Schwarzschild singularity},''
  \href{http://dx.doi.org/10.1103/PhysRev.108.1063}{{\em Phys. Rev.} {\bf 108}
  (1957)  1063--1069}.

\bibitem{Martel:2005ir}
K.~Martel and E.~Poisson, ``{Gravitational perturbations of the Schwarzschild
  spacetime: A Practical covariant and gauge-invariant formalism},''
  \href{http://dx.doi.org/10.1103/PhysRevD.71.104003}{{\em Phys.\ Rev.\ D} {\bf
  71} (2005)  104003}, \href{http://arxiv.org/abs/gr-qc/0502028}{{\tt
  arXiv:gr-qc/0502028}}.

\bibitem{TheLIGOScientific:2017qsa}
{\bf LIGO Scientific, Virgo} Collaboration, B.~Abbott {\em et al.},
  ``{GW170817: Observation of Gravitational Waves from a Binary Neutron Star
  Inspiral},'' \href{http://dx.doi.org/10.1103/PhysRevLett.119.161101}{{\em
  Phys. Rev. Lett.} {\bf 119} (2017) no.~16, 161101},
  \href{http://arxiv.org/abs/1710.05832}{{\tt arXiv:1710.05832 [gr-qc]}}.

\bibitem{GBM:2017lvd}
{\bf LIGO Scientific, Virgo, Fermi GBM, INTEGRAL, IceCube, AstroSat Cadmium
  Zinc Telluride Imager Team, IPN, Insight-Hxmt, ANTARES, Swift, AGILE Team,
  1M2H Team, Dark Energy Camera GW-EM, DES, DLT40, GRAWITA, Fermi-LAT, ATCA,
  ASKAP, Las Cumbres Observatory Group, OzGrav, DWF (Deeper Wider Faster
  Program), AST3, CAASTRO, VINROUGE, MASTER, J-GEM, GROWTH, JAGWAR,
  CaltechNRAO, TTU-NRAO, NuSTAR, Pan-STARRS, MAXI Team, TZAC Consortium, KU,
  Nordic Optical Telescope, ePESSTO, GROND, Texas Tech University, SALT Group,
  TOROS, BOOTES, MWA, CALET, IKI-GW Follow-up, H.E.S.S., LOFAR, LWA, HAWC,
  Pierre Auger, ALMA, Euro VLBI Team, Pi of Sky, Chandra Team at McGill
  University, DFN, ATLAS Telescopes, High Time Resolution Universe Survey,
  RIMAS, RATIR, SKA South Africa/MeerKAT} Collaboration, B.~Abbott {\em et
  al.}, ``{Multi-messenger Observations of a Binary Neutron Star Merger},''
  \href{http://dx.doi.org/10.3847/2041-8213/aa91c9}{{\em Astrophys. J. Lett.}
  {\bf 848} (2017) no.~2, L12}, \href{http://arxiv.org/abs/1710.05833}{{\tt
  arXiv:1710.05833 [astro-ph.HE]}}.

\bibitem{Monitor:2017mdv}
{\bf LIGO Scientific, Virgo, Fermi-GBM, INTEGRAL} Collaboration, B.~Abbott {\em
  et al.}, ``{Gravitational Waves and Gamma-rays from a Binary Neutron Star
  Merger: GW170817 and GRB 170817A},''
  \href{http://dx.doi.org/10.3847/2041-8213/aa920c}{{\em Astrophys. J. Lett.}
  {\bf 848} (2017) no.~2, L13}, \href{http://arxiv.org/abs/1710.05834}{{\tt
  arXiv:1710.05834 [astro-ph.HE]}}.

\bibitem{Creminelli:2016zwa}
P.~Creminelli, D.~Pirtskhalava, L.~Santoni, and E.~Trincherini, ``{Stability of
  Geodesically Complete Cosmologies},''
  \href{http://dx.doi.org/10.1088/1475-7516/2016/11/047}{{\em JCAP} {\bf 11}
  (2016)  047}, \href{http://arxiv.org/abs/1610.04207}{{\tt arXiv:1610.04207
  [hep-th]}}.

\bibitem{Solomon:2017nlh}
A.~R. Solomon and M.~Trodden, ``{Higher-derivative operators and effective
  field theory for general scalar-tensor theories},''
  \href{http://dx.doi.org/10.1088/1475-7516/2018/02/031}{{\em JCAP} {\bf 02}
  (2018)  031}, \href{http://arxiv.org/abs/1709.09695}{{\tt arXiv:1709.09695
  [hep-th]}}.

\bibitem{Franciolini:2018aad}
G.~Franciolini, L.~Hui, R.~Penco, L.~Santoni, and E.~Trincherini, ``{Stable
  wormholes in scalar-tensor theories},''
  \href{http://dx.doi.org/10.1007/JHEP01(2019)221}{{\em JHEP} {\bf 01} (2019)
  221}, \href{http://arxiv.org/abs/1811.05481}{{\tt arXiv:1811.05481
  [hep-th]}}.

\end{thebibliography}\endgroup

\end{document}